\documentclass[10pt,letterpaper]{article}
\usepackage[top=0.85in,left=0.8in,footskip=0.75in,marginparwidth=0.5in]{geometry}

\usepackage[utf8]{inputenc}

\usepackage{cite}

\usepackage{microtype}
\DisableLigatures[f]{encoding = *, family = * }

\usepackage{changepage}

\usepackage[aboveskip=1pt,labelfont=bf,labelsep=period,singlelinecheck=off]{caption}

\makeatletter
\renewcommand{\@biblabel}[1]{\quad#1.}
\makeatother

\usepackage{lastpage,fancyhdr,graphicx}
\usepackage{epstopdf}
\fancyheadoffset[L]{2.25in}
\fancyfootoffset[L]{2.25in}

\usepackage{color}

\definecolor{Gray}{gray}{.25}

\usepackage{graphicx}

\usepackage{wrapfig}
\usepackage[pscoord]{eso-pic}
\usepackage[fulladjust]{marginnote}
\reversemarginpar

\usepackage{subfigure}
\usepackage{siunitx}
\usepackage{gensymb}
\usepackage{epstopdf}
\usepackage{url}

\usepackage{amsmath}
\usepackage{amssymb}
\usepackage{amsfonts}

\newcommand{\bb}{\boldsymbol}

\def\mysingleq#1{`#1'}

\begin{document}
\vspace*{0.35in}

% title goes here:
\begin{flushleft}
{\Large
\textbf\newline{Classes of low-frequency earthquakes based on inter-time distribution reveal a precursor event for the 2011 Great Tohoku Earthquake}
}
\newline
% authors go here:
\\
Tomoki Tokuda \textsuperscript{*} \&
Hirohiko Shimada \textsuperscript{}
\\
\bigskip
\bf{} Okinawa Institute of Science and Technology Graduate University,
             1919-1 Tancha, Okinawa, 904-0495, JAPAN
\\
\bigskip
* tomoki.tokuda@oist.jp

\end{flushleft}

\section*{Abstract}
Recently, slow earthquakes (slow EQ) have received much attention relative to understanding the  mechanisms underlying large earthquakes and to detecting their precursors. Low-frequency earthquakes (LFE) are a specific type of slow EQ. In the present paper, we reveal the relevance of LFEs to the 2011 Great Tohoku Earthquake (Tohoku-oki EQ) by means of cluster analysis. We classified LFEs in northern Japan in a data-driven manner, based on inter-time, the time interval between neighboring LFEs occurring within 10 km. We found that there are four classes of LFE that are characterized by median inter-times of 24 seconds, 27 minutes, 2.0 days, and 35 days, respectively. Remarkably, in examining the relevance of these classes to the Tohoku-oki EQ, we found that activity in the shortest inter-time class (median 23 seconds) diminished significantly at least three months before the Tohoku-oki EQ, and became completely quiescent 30 days before the event (p-value = 0.00014). Further statistical analysis implies that this class, together with a similar class of volcanic tremor, may have served as a precursor of the Tohoku-oki EQ. We discuss a generative model for these classes of LFE, in which the shortest inter-time class is characterized by a generalized gamma distribution with the product of shape parameters 1.54 in the domain of inter-time close to zero. We give a possible geodetic interpretation for the relevance of LFE to the Tohoku-oki EQ.

% now start line numbers
%\linenumbers

% the * after section prevents numbering

\section*{Introduction}
Detecting and identifying precursors to large earthquakes is essential in order to mitigate the devastating damage of earthquakes. In relation to the 2011 Great Tohoku Earthquake (Tohoku-oki EQ, hereafter) $M_w$ 9.0, several precursors were reported retrospectively.  With a long-term perspective, seismic quiescence was observed near the epicenter 23 years prior to Tohoku-oki EQ \cite{katsumata2011long}. Also, strong correlations were observed between tidally induced stresses and earthquake occurrence times near the epicenter ten years before it occurred \cite{tanaka2012tidal}. Further, an anomaly in the $b$-value of the Gutenberg-Richter law was reported near the epicenter five years before \cite{nanjo2012decade}. In the short term, it was observed that earthquake activities moved toward the epicenter one month before \cite{kato2012propagation}. Regarding  non-seismic phenomena, positive anomalies of ionospheric total electron content (TEC) were detected around the focal region 40 minutes before \cite{heki2011ionospheric}. Further, anomalous changes of groundwater levels and Radon concentrations were reported three months before \cite{orihara2014preseismic, tsunomori2014anomalous}.
% * <steven.aird@oist.jp> 2018-07-10T03:30:45.717Z:
% 
% > law
% Tomoki, usually it would be incorrect to refer to an anomaly in a law.   It would be an anomaly in some sort of statistic.
% 
% ^ <steven.aird@oist.jp> 2018-07-10T03:32:00.263Z.
All these precursors are important not only for prediction of  large earthquakes, but also for a better understanding of their underlying mechanisms.  In the present study, we examine the relevance of slow earthquakes (slow EQ) to prediction of major events, focusing in particular on low-frequency earthquakes (LFE).

Slow EQs are low-frequency phenomena, distinguished from regular earthquakes, with lower dominant frequencies ranging from several Hz to several years \cite{obara2016connecting}. Slow EQs include several subtypes such as 
LFEs, very low-frequency earthquakes, short-term slow-slip events, and long-term slow-slip events, depending on the range of dominant frequencies. Recently, such earthquakes have gained much attention both for better understanding the underlying mechanisms of earthquakes and for identifying precursors of large earthquakes \cite{obara2002nonvolcanic, peng2010integrated, shelly2007non}. Indeed, anomalous occurrences of slow EQs have been reported prior to large earthquakes \cite{roeloffs2006evidence}. In case of the Tohoku-oki EQ, it is inferred that slow-slip events occurred over a period of one month prior to the megathrust EQ \cite{ito2013episodic}. It is implied that these slow events increased shear stress across a wide swath of the Tohoku-oki region, which eventually triggered the Tohoku-oki EQ. In the Parkfield EQ,  an anomalous change in occurrences of slow tremors was observed near the hypocenter three months before the earthquake \cite{shelly2009possible}. 

Slow EQs are commonly categorized into two categories:  aseismic (geodetic) events and seismic events. The former include slow-slip events during short or long periods of time, while the latter include single low-frequency earthquakes (LFEs) and tremors comprising a number of LFEs. Despite various manifestations, slow EQs are basically caused by shear slips, the same as regular earthquakes \cite{ide2007scaling}. However, slow EQs follow a different scaling law from regular earthquakes in seismic moment and characteristic duration. This suggests a longer duration of slow EQs for a given seismic moment than conventional earthquakes, though their mechanism is not well understood. The association between slow EQs and large earthquakes may be that slow EQs  work as stress meters both in the long- and short-term.  On one hand, long-term, slow-slip events reflect dominant changes in strains, which enable one to evaluate precise accumulation of strains \cite{obara2016connecting}. On the other hand,  short-term, slow-slip events or tremors may reflect nucleation processes leading to large earthquakes \cite{ohnaka1992earthquake, kato2012propagation}. These assumptions suggest the importance of slow EQs as precursors for large EQs, both long- and short-term.

Some of the most compelling studies have focused on slow EQs near the epicenters of major earthquakes \cite{ito2013episodic, kato2012propagation}. In particular, it was suggested that a nucleation process (that accelerates slow-slip movement, finally triggering a large earthquake) in the form of slow-slip events occurring near the epicenter.  However, if slow EQs work as a true stress indicator, we would expect that some slow EQ anomaly might also occur even distant from the epicenter, because changes of stress prior to a large EQ may extend across a wide area of a continental plate. For instance, anomalous strain changes were identified in a borehole located in the Oshika Peninsula 150 km away from the epicenter before the Tohoku-oki EQ \cite{ito2013episodic}. Furthermore, we would expect that such anomalous slow EQs, which occur away from the epicenter along the downdip edge of the megathrust seismogenic zone, might take the form of LFEs \cite{obara2016connecting}. To the best of our knowledge, however, there has been no study that tried to shed light on LFEs far from the epicenter of a large EQ. The overall goal of the present study is to examine behavior of LFEs away from the epicenter of the Tohoku-oki EQ. 

\begin{figure}[ht!]
     \begin{center}
     \includegraphics[scale=0.4, trim=60mm 0mm 0mm 0mm]{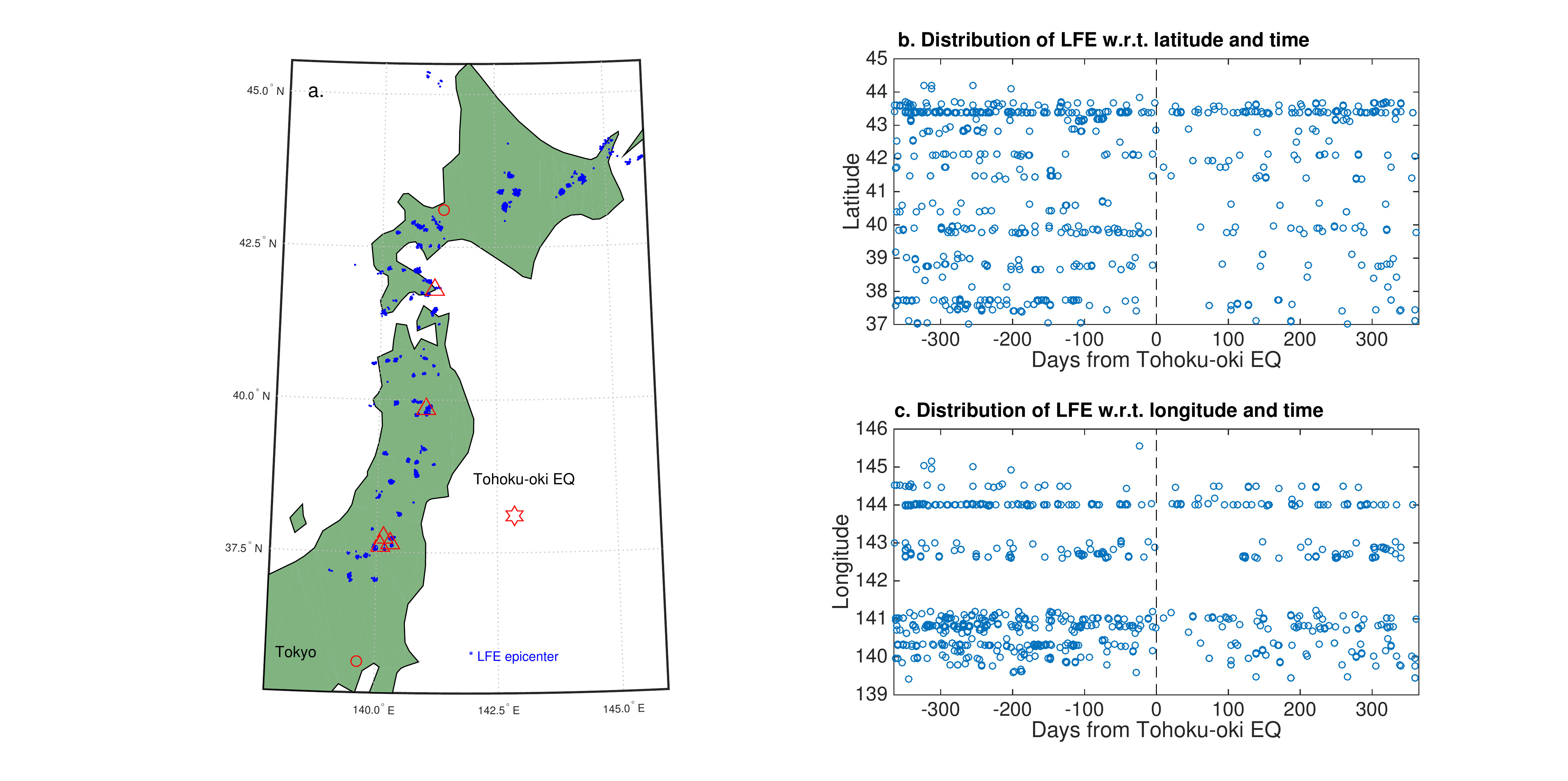}  
                \end{center}
    \vspace{0mm}
        \caption{Distribution of LFE. Panel a. Latitude and longitude distribution. 
        Locations of epicenters of LFE (blue dots) and 
        five volcanoes (triangles):
        Esan, Iwatesan, Azumayama, Adatarayama, and Bandaisan
       from North to South. The red hexagon
        denotes the epicenter of the Tohoku-oki EQ. Panel b. Latitude and temporal distribution. 
        Panel c. Longitude and temporal distribution.}
          \label{map}
\end{figure}

Toward this end, we studied LFEs that occurred in northern Japan, several hundred kilometers from the epicenter of the Tohoku-oki EQ (Fig.\ref{map}a). The underlying physical mechanism of LFE in this region is not fully understood \cite{hasegawa1994deep, katsumata2003low}. However, it is inferred from the analysis of velocities of P- and S-waves that LFEs in this region are related to aqueous fluids supplied by the subducted slab \cite{hasegawa2005deep}. These fluids, originating from the mantle wedge in the continental plate, move up, resulting in accumulation of a large amount of melt below the Moho discontinuity. It is speculated that a sudden movement of such fluids near the Moho discontinuity may cause LFEs. Further, it is observed that the low seismicity of LFEs in this region seems synchronized with that of earthquakes in the Wadati-Benioff zone or the shallow inland seismicity, which implies that generation of LFEs may be related to changes in tectonic stress fields over a wide area \cite{hasegawa1994deep}. With regard to the Tohoku-oki EQ, it is reported that in a spatial analysis, LFEs became generally less active after the Tohoku-oki EQ\cite{Kosuga2017}, which contrasts with activation of conventional EQs in this region (Fig.\ref{map}b, c). However, such a spatial analysis does not reveal whether such a change of activation occurred before the Tohoku-oki EQ. 
% * <steven.aird@oist.jp> 2018-07-10T04:57:28.463Z:
% 
% > melts below the Moho
% Tomoki, I do not understand this.
% 
% ^ <steven.aird@oist.jp> 2018-07-10T04:57:45.255Z.
% * <steven.aird@oist.jp> 2018-07-10T04:19:55.160Z:
% 
% > melts
% Melts of what?  What is the Moho?
% 
% ^ <steven.aird@oist.jp> 2018-07-10T04:20:09.069Z.

In the present paper, to clarify this point, we employ a different approach to analyze LFEs, focusing on the inter-time distribution, the time between consecutive events, rather than on their spatial distribution.  For conventional EQs, the distribution of inter-time has recently attracted attention because the inter-time with a cutoff magnitude (from 5 to 6.5) seems to entail a universal law in the form of a generalized gamma distribution\cite{bak2002unified, corral2004long, molchan2005interevent, de2016statistical}. With respect to the Tohoku-oki EQ, the parameters of such a generalized gamma distribution in the Tohoku region changed (as seen later in Table~\ref{gampareto}). In the context of LFEs, it is not obvious what distribution the inter-time follows. Nonetheless, it is of great interest to examine whether the distribution of inter-time changed with respect to the Tohoku-oki EQ. If we can identify exactly when the change of distribution occurred, this provides useful information on a possible precursor for the Tohoku-oki EQ. To shed light on this, we focused on the inter-time of LFEs that occurred within 10 km of the epicenter, which captures correlations of consecutive LFEs.  To identify homogeneous distributions, we performed cluster analyses on logarithms of LFE inter-time in a data-driven manner, which identified four homogeneous classes. Remarkably, examination of the relevance of these classes to the Tohoku-oki EQ suggests that the activity of LFEs in the shortest inter-time class (median 24 seconds) diminished significantly at least three months before the Tohoku-oki EQ, and that complete quiescence occurred in this class 30 days before the Tohoku-oki EQ (p-value = 0.00014). In contrast with LFEs, conventional earthquakes did not become inactive during the same period. Further statistical analysis implies that this class together with a similar class of volcanic tremor may have served as a precursor of the Tohoku-oki EQ. Here, we refer to \mysingleq{precursor} as a phenomenon that does not cause a large earthquake, but serves as a useful indicator. We discuss a generative model in terms of non-homogenous Poisson processes for these classes of LFE in which the shortest inter-time class is characterized by a generalized gamma distribution with 
a product of shape parameters 1.54 in the domain of inter-time close to zero. Lastly, we give a possible geodetic interpretation for the relevance of LFEs to the Tohoku-oki EQ.  
% * <steven.aird@oist.jp> 2018-07-10T05:06:39.821Z:
% 
% > precursor
% Tomoki, are you suggesting that the LFEs actually caused the Tohoku EQ or that they simply served as indicators of it?
% 
% ^ <steven.aird@oist.jp> 2018-07-10T05:07:28.913Z.
% * <steven.aird@oist.jp> 2018-07-10T04:59:41.260Z:
% 
% > convective
% Tomoki, please check this word.
% 
% ^ <steven.aird@oist.jp> 2018-07-10T05:00:12.629Z.

\section*{Results}
We considered LFEs that occurred along the volcanic front in northern Japan (latitude greater than 
$37\degree$N; Fig.\ref{map}a). Some of their locations are close to active volcanoes, while others are not.  A close relationship between LFEs and volcanic activity is anticipated, but is not clearly understood \cite{Takahashideep2009}.
We pre-processed LFE data to obtain inter-times between consecutive events. We considered two types of datasets, taking into account proximity of consecutive LFEs. First, we evaluated inter-times of remote LFEs that are separated more than 10 km. Hereafter, we refer to this dataset as \mysingleq{Remote LFE}. Second, we evaluate inter-times of neighboring LFEs less than 10 km, referring to this dataset as \mysingleq{Neighbour LFE} (see more details in data and pre-processing in section of Methods). 

For purposes of comparison, we also included volcanic tremors and conventional earthquakes in our analysis.  We considered volcanic tremors that occurred in five volcanoes along the volcanic front (between $37\degree$-$41\degree$N, Fig.~\ref{map}a) with sufficient inter-time sample size. First, we evaluated the inter-time between two volcanic tremors that occurred in different volcanoes. We define this dataset as \mysingleq{Remote volcanic tremors}. Second, we separately evaluated the inter-time for each volcano, which has the same effect as constraining the proximity between two events, just as among neighboring pairs of LFEs (\mysingleq{Neighbour volcanic tremors}). Note that we did not take into account possible overlaps of samples of inter-times between LFEs and volcanic tremors, which are negligible \footnote{However, this does not preclude interplay between LFEs and other volcanic phenomena, like tremors and EQ swarms. For instance, it is worth noting that more than 95\% epicenters of class S2 during the bursting period [-950, -900] days (Fig. \ref{fig4}a) were actually localized inside a narrow square region of latitudes 43.34\degree N-43.44\degree N and longitudes 143.95\degree E-144.05\degree E in the vicinity of the Meakandake volcano. This burst preceded the singular tremors \cite{ogiso2012migration} that started on day -893 (September 29, 2008) by $7$ days. Further analysis of the characterization and intermittency of S2, based on such observations would be interesting.}. We also included conventional earthquakes in our analysis, for which we used the same definition of \mysingleq{remote} and \mysingleq{neighbor} as in LFE. We did not restrict conventional EQs based upon magnitude.

We transformed these datasets into logarithms of base 10, which were subsequently analyzed. For \mysingleq{neighbour} datasets, we carried out cluster analyses to reveal the underlying stochastic model to generate data (see more details about cluster analyses in that section of the Methods). 

\subsubsection*{LFE}
\begin{figure}[]
\begin{flushleft}
\includegraphics[scale=0.4, trim=30mm 0mm 0mm 0mm]{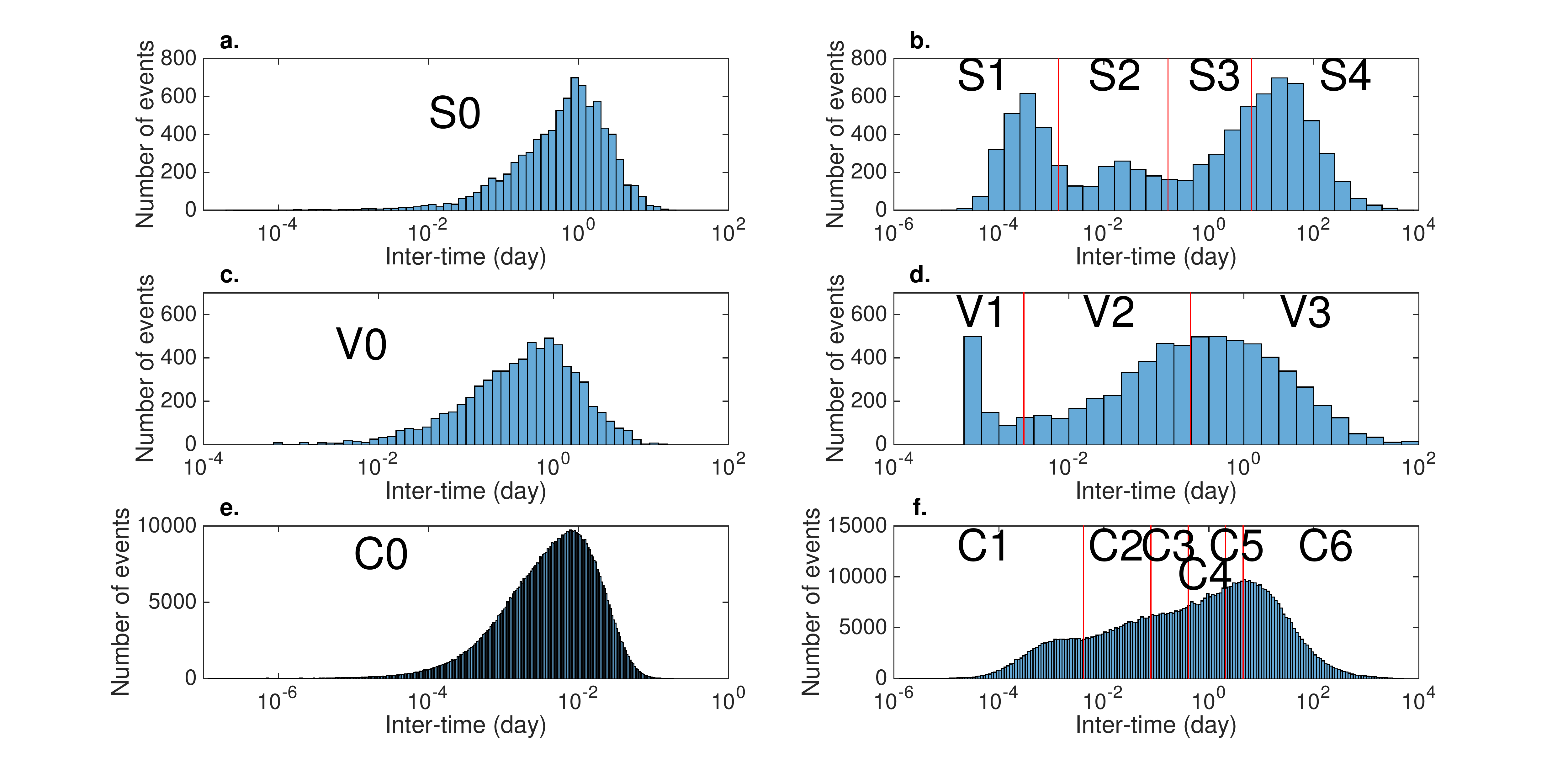}  
\end{flushleft}
\caption{Distribution of inter-times between earthquakes. 
Panel a: Inter-time distribution of remote pairs of LFEs.
Panel b: Inter-time distribution of neighboring pairs of LFEs. Red lines denote borders between four classes S1, S2, S3, and S4, which were yielded by fitting Gaussian mixture models to the logarithm of inter-time. 
Panel c: Inter-time distribution of remote pairs of volcanic tremors. Here, we define \mysingleq{remote} as different volcanoes.  Panel d: Inter-time distribution of neighboring volcanic tremors, where 
the definition \mysingleq{neighboring} denotes same volcano. Red lines denote borders between three classes V1, V2 and V3. Panel e and f: Inter-time distribution for conventional EQ.}
\label{histint}
\end{figure}

For remote LFEs, the inter-time ranges widely from 2.16 seconds to 17.1 days.  It appears that the distribution of the logarithm of inter-time is unimodal with small skewness (Fig.~\ref{histint}a). Since the inter-distance between consecutive events in this dataset is more than 10 km, it is reasonable to assume that the direct causal relationship between consecutive events is relatively small. We did not carry out further analysis for this dataset, but keep it as a reference distribution (denoting this class of inter-time as S0, Table~\ref{GaussEstimation}). For neighboring LFEs, it appears that the distribution of the logarithm of inter-time is multimodal, with three modes (Fig.~\ref{histint}b), which could suggest at least three underlying distributions for this dataset. Fitting Gaussian mixture models suggests four clusters. The estimates of means and variances for each Gaussian component are summarized in Table~\ref{GaussEstimation}. Following a conventional procedure of mixture models, we clustered the logarithms of inter-time,  yielding four classes. Here, classes S1-S4 are arranged in ascending order of means. Note that the optimal mixture model identified two distinct classes, S3 and S4, which initially appeared to comprise a single cluster. Notably, classes S2-S4 have similar variances close to 0.40, whereas class S1 has a considerably smaller variance, 0.17. One may characterize the inter-time scale of these classes as follows: seconds for S1, minutes for S2, days for S3, and months for S4 (column of \mysingleq{Median Inter-time} in Table~\ref{GaussEstimation}). 

\subsubsection*{Volcanic tremors}
For remote volcanic tremors, the inter-time  ranges widely from 60.0 seconds to 17.4 days.  It appears that the distribution of the logarithm of inter-time is unimodal with small skewness (Fig.~\ref{histint}c). For neighboring volcanic tremors, we aimed to identify distinct classes of inter-time in a similar manner as with LFEs. First, we extracted samples with very short inter-times, which do not seem to follow a Gaussian distribution (V1). For the remaining samples, we applied Gaussian mixture models, which identified four optimal classes. Among these four, we combined two classes that comprise less than $5\%$ of those samples. As a consequence, we had three classes for inter-times of volcanic tremors V1, V2 and V3 (Fig.~\ref{histint}d, Table~\ref{GaussEstimation}). As with LFEs, one may characterize inter-time scales of these classes as minutes for V1, hours for V2, and days for V3 (Table~\ref{GaussEstimation}).

\subsubsection*{Conventional EQ}
For remote conventional EQs, inter-time ranges from 0.01 seconds to 4.26 hours.  It appears that the distribution of the logarithm of inter-time  is unimodal with small skewness (Fig.~\ref{histint}e). For neighboring conventional EQs,  we carried out a cluster analysis to identify six clusters (Fig.~\ref{histint}f, Table~\ref{GaussEstimation}).
In contrast with the results for LFEs, the variance in each cluster varies considerably. Interestingly, however, the variance of class C0 has a similar value to that of class S0 (0.402 and 0.399, respectively). We will return to this point in the Discussion. 

\subsubsection*{Comparison of distributions}
% * <steven.aird@oist.jp> 2018-07-10T06:01:14.167Z:
% 
% > Comparisons
% Comparisons of what?
% 
% ^ <steven.aird@oist.jp> 2018-07-10T06:01:24.524Z.
Distributions of inter-time have been empirically/theoretically studied for conventional EQs \cite{bak2002unified, saichev2007theory, corral2004universal, lippiello2013magnitude}, which suggest that its shape approximates a gamma distribution, but the exact distribution is not known. Currently, there is no consensus on the exact distribution of intertimes. In the present paper, we have fitted Gaussian mixture models  to log-tramsformed inter-times. From the point of view of inter-time, this amounts to fitting a log-normal distribution to each class:
\begin{eqnarray}
   f(x) \sim \frac{1}{x \sqrt{\sigma^2}}\exp \big \{-\frac{(\log_{10} x - \mu)^2}{2\sigma^2}\big \},
   \label{lognormal}
\end{eqnarray}
where $x$ is inter-time, and $\mu$ and $\sigma^2$ are the mean and variance of log-transformed inter-time.
As can be seen in Eq.~(\ref{lognormal}),  the value of $\sigma^2$ is scale-invariant with respect to data: through the transformation of $x \rightarrow a x$,  the variance $\sigma^2$ does not change. Hence, the characterization of a distribution by variance $\sigma^2$ provides a useful tool to compare the shape of the distribution for data of different time-scales. Remarkably, the classes of LFE and volcanic tremors in our data are characterized by two specific values of $\sigma^2$ (Table~\ref{GaussEstimation}): Classes S0, S2, S3, S4, V0, and V3 have variances $\sigma^2$ close to 0.4,  whereas classes S1 and V2 have variance close to 0.2. This observation suggests that these two groups may have different underlying mechanisms of occurrence.
    
\subsection*{Anomalies related to the Tohoku-oki EQ}
Once our data-driven classification was performed, certain anomalous seismicities of LFEs and volcanic tremors became evident with respect to the timing of the Tohoku-oki EQ (Fig.~\ref{evolraw}). It appears that the seismicities of some classes of LFE and volcanic tremors changed with respect to the Tohoku-oki EQ. Importantly, such a change in S1 and V1 seems to occur just before the Tohoku-oki EQ (red circle in Fig.~\ref{evolraw}). We investigated such anomalies, focusing on seismicity in each class from both long- and short-term perspectives. Here, long-term signifies several months while short-term indicates several days. For simplicity, we transformed the time of occurrence as $t'=t-t_{tohoku}$ where $t$ is the time of occurrence and $t_{tohoku}$ is the origin time of the Tohoku-oki EQ. 

\subsubsection*{Long-term anomaly}
We examined a long-term anomaly, focusing on the evolution of seismicity in each class of LFE (Fig.~\ref{fig4}a, c).  A brief inspection of Fig.~\ref{fig4}c suggests that the rate of occurrences decreased after the Tohoku-oki EQ 
for classes S1-S4. In particular, such a change was drastic for class S1, which can be observed as a \mysingleq{kink} in the evolution of cumulative number of occurrence (Fig.~\ref{fig4}a). Moreover, the decrease in rate of occurrence seems to have occurred some days before the Tohoku-oki EQ (Fig.~\ref{fig4}c). We examined this hypothesis by means of ROC (Receiver Operating Characteristic) analysis \cite{fawcett2006introduction}, which has been widely used for decision-making in the domains of medical research and machine learning community (see section of ROC analysis in the Methods). 
To apply this analysis to our context, we manipulated the cutoff day from -300 to 300 days (we assume the cutoff day takes only integers), which splits the data points into two groups: a before-group (the time of occurrence is less than the cutoff day) and an after-group (the time of occurrence is larger than the cutoff day). This procedure yielded a binary label for each data point, showing the group to which the data point  belonged. Using this labeling for each cutoff day, we evaluated the area under the curve (AUC) of the number of occurrences based on a logistic regression analysis. Finally, we evaluated the maximum AUC and the corresponding cutoff day (Table~\ref{GaussEstimation}). 

We found that class S1 gives the largest value of AUC (0.83), followed by class S3 (0.72),  class S4 (0.70), and class S2 (0.44). This suggests that the rate of occurrence changed most in S1. Remarkably, the optimal cutoff day for this class is -76 days, i.e., 76 days before the Tohoku-oki EQ, which quantitatively suggests that an anomalous behavior (quiescence) occurred 2.5 months before the Tohoku-oki EQ.  Similar analysis was performed for volcanic tremors (Fig.~\ref{fig4}b, Table~\ref{GaussEstimation}). In this case, however, the optimal cutoff day is positive, not capturing a long-term anomaly prior to the Tohoku-oki EQ. 

The question then arose whether quiescence of class S1 could be detected if we use only data before the Tohoku-oki EQ. In terms of prediction, it is important to capture such a precursor before the earthquake. For this purpose, we evaluated \mysingleq{Z-value}  \cite{wiemer2000minimum, katsumata2011long}, which is useful for measuring seismic quiescence: A large value of Z provides evidence of seismic quiescence during a target period. Note that we did not consider spatial differences of seismicity in this analysis, but we used all data irrespective of the epicenters. The results of Z-values are displayed in Fig.~\ref{fig4}d in which class S1 clearly shows quiescence approaching the Tohoku-oki EQ, while the remaining classes did not. In the beginning of 2010, the Z-value of class S1 was low, but it jumped over 2 when the window focuses on the target period between -229 days and -109 days (the width of window is 120 days), which shows that the Z-value analysis detected quiescence of class S1 earlier than in the AUC analysis (-76 days), i.e., at least three months before the Tohoku-oki EQ. After this period, the Z-value of class S1 remained as high as 4 just before the Tohoku-oki EQ.  On the other hand, the Z-values of S2, S3 and S4 were not high, just before the Tohoku-oki EQ, which suggests that these classes do not show quiescence near the Tohoku-oki EQ.

\subsubsection*{Short-term anomaly} 
Next, we investigated the behavior of seismicity in more detail from a short-term perspective. It is of great interest to detect short-term anomalies, which might work as an immediate portent of the Tohoku-oki EQ. 

\begin{figure}[]
     \begin{flushleft}
    \includegraphics[scale=0.4, trim=50mm 0mm 0mm 0mm]{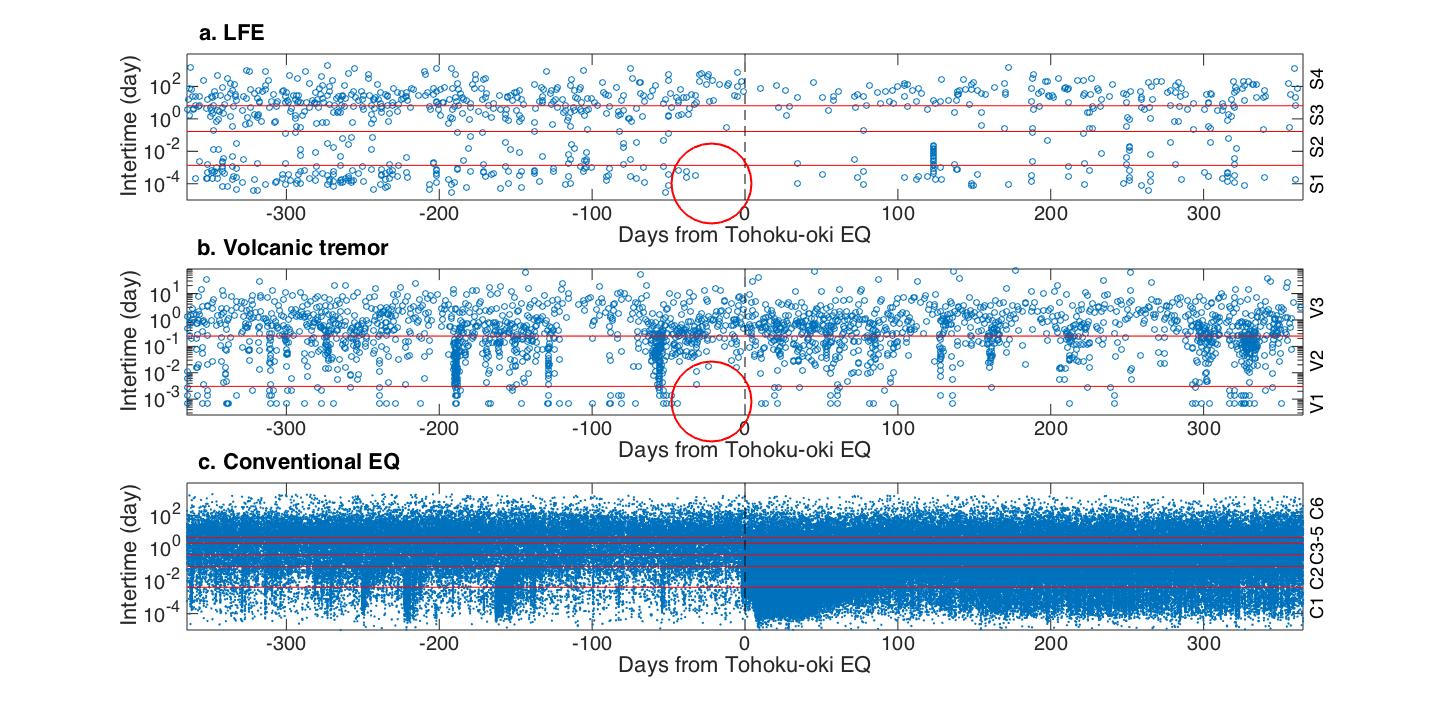} 
                \end{flushleft}
    \vspace{0mm}
        \caption{Raw data showing occurrence of neighboring events one year before and after the Tohoku-oki EQ. 
         Panel a: LFE.  Each point denotes the pre-event of a pair of LFEs for a given inter-time. The
        x-axis denotes days from the Tohoku-oki EQ while the
        y-axis denotes inter-time. The red lines indicate borders between classes.
        LFEs of class S1 ceased before the Tohoku-oki EQ (red circle). Panels b and c: the counterparts of Panel a for volcanic tremors and conventional EQs. }
        \label{evolraw}
\end{figure}

\begin{figure}[]
     \begin{center}
     \includegraphics[scale=0.4, trim=40mm 0mm 0mm 0mm]{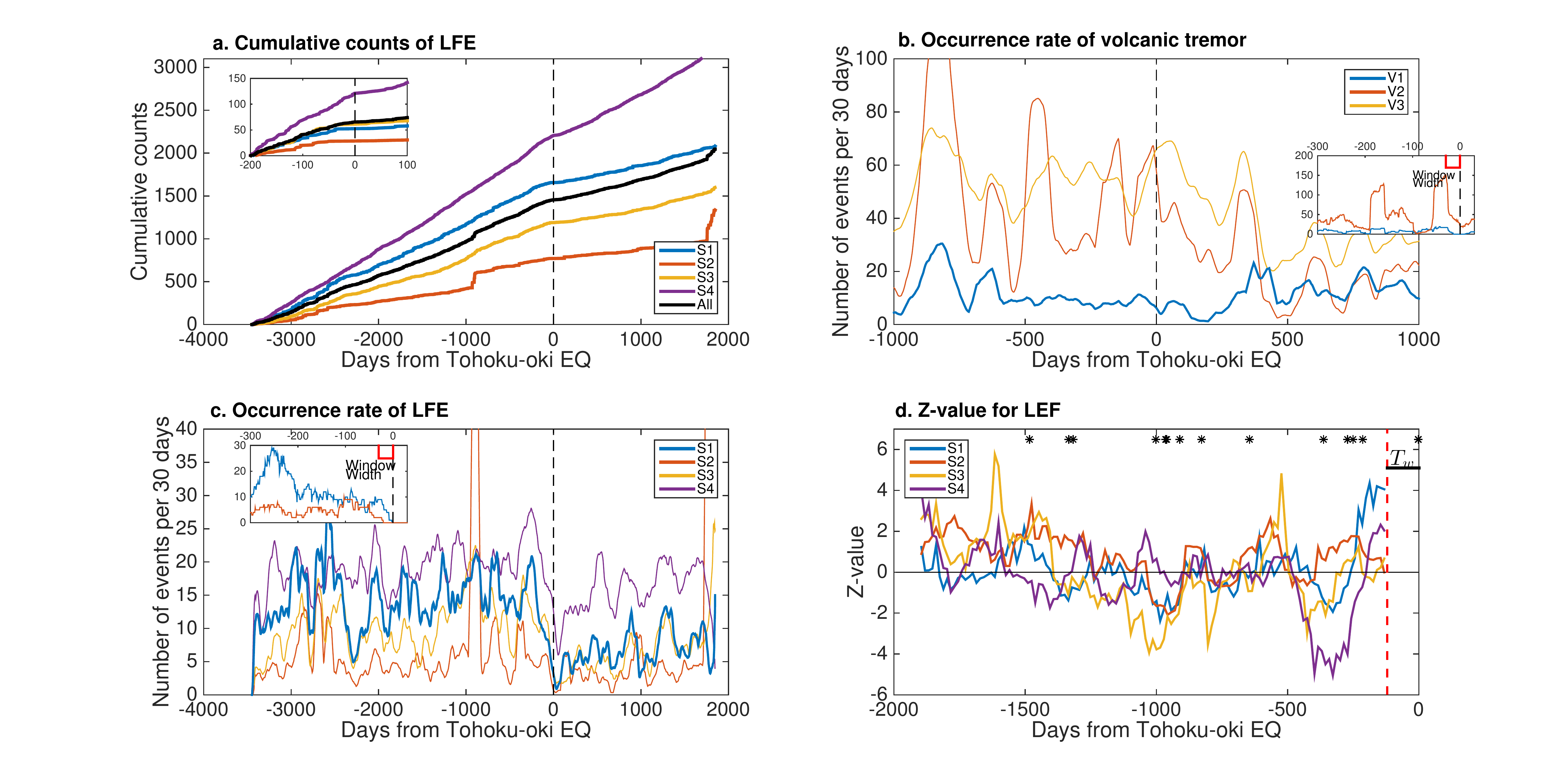}  
                \end{center}
    \vspace{0mm}
        \caption{Processed data regarding the occurrence of neighboring events. Panel a: 
        Evolution of cumulative LFEs for classes S1, S2, S3 and S4. The black curve denotes  cumulative events for all classes divided by four. Note that 
        the bursting behavior of S2 is observed in the period between -950 and -900 days and in the period between
         1750 and 1850 days. In these periods, more than 95\% of epicenters of S2 are localized
         in the square area between
         latitudes 43.34\degree N-43.44\degree N and longitudes 143.95\degree E-144.05\degree E close to the Meakandake volcano.
         Panel b:  Cumulative volcanic tremors for classes V1, V2 and V3. The number of volcanic tremor is evaluated in the same manner as in Panel c. 
         Panel c: Evolution of LFEs  for classes S1, S2, S3 and S4. The number of LFEs
        denoted by $n(z)$ is evaluated with the width of window set 30 days backward:  
        $n(z) = \sum_{i=1}^{N} \mathbb{I}(t_i' \geq  z-30)\mathbb{I}(t_i' \leq z) $ where $\mathbb{I}(\cdot)$ is an
        indicator function; $t_i' = t_i - t_{tohoku}$ with $t_{tohoku}$ being the origin time of the Tohoku-oki EQ; $z$ is an integer
        ranging from -3000 to 2000. 
      For the main graph, $n(z)$ is further smoothed using a moving average of a 90 day-window (45 days backward, 45 days forward)  while it is not smoothed in the inset. 
      Panel d: Z-values for each class of LFE. We set the width of window ($T_w$) to 120 days and 
        the moving step to 14 days. The initial time ($t_0$) is Jan. 1, 2006, while the terminal time ($t_e$) is 
        March 8, 2011. We did not consider spatial differences, but instead we used all LFE data  for each class.
        On the top of the panel, the occurrence of large earthquakes with magnitude larger than 6 is denoted by 
        black asterisks. 
         }
        \label{fig4}
\end{figure}

\begin{figure}[]

     \begin{flushleft}
     \includegraphics[scale=0.4, trim=50mm 0mm 0mm 0mm]{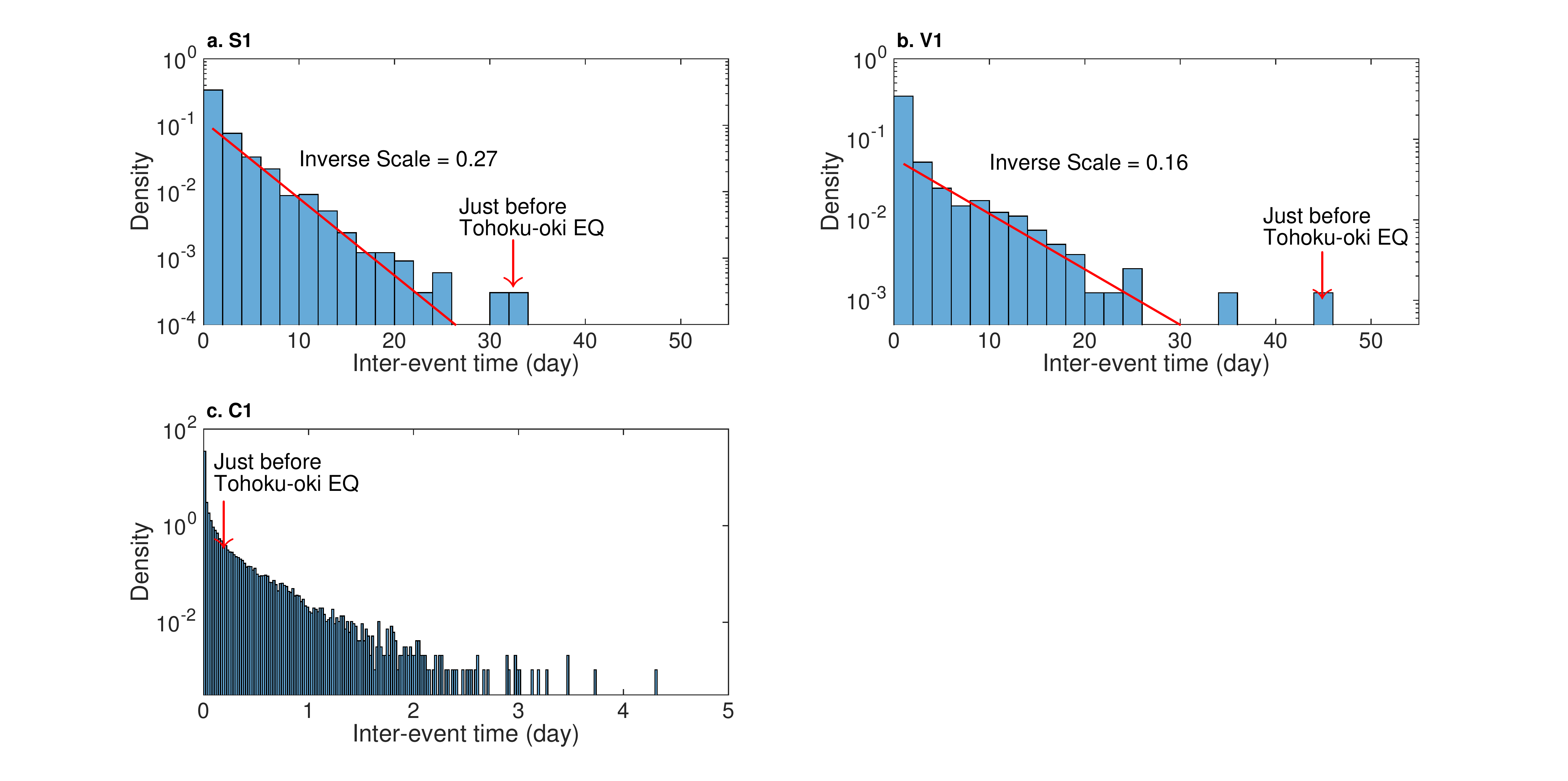} 
                \end{flushleft}
        \caption{ Distributions of inter-event time (density). Panel a for class S1. 
        In this panel, we discarded inter-event times smaller than the cutoff interval for class S1. 
               The red line was 
     estimated by fitting a truncated exponential distribution with lower cutoff of 1 day where the mean was evaluated by a robust statistics with median/log(2), which is less influenced by outliers \cite{gather1999robust}. The estimated density function was normalized such that the sum of probability that inter-event time is greater than one day matches the empirical one. 
% * <steven.aird@oist.jp> 2018-07-10T08:13:48.492Z:
% 
% > one
% "One" what?
% 
% ^ <steven.aird@oist.jp> 2018-07-10T08:13:59.794Z.
% * <steven.aird@oist.jp> 2018-07-10T08:12:56.860Z:
% 
% >  robust statistics 
% Tomoki, I do not know what this means.
% 
% ^ <steven.aird@oist.jp> 2018-07-10T08:13:10.265Z.
     The slope of the line (inverse scale for exponential distribution) is displayed as text. 
   Panel b: the counterpart of Panel a for class V1 of volcanic tremors.  Panel c: the counterpart for class C1 for conventional EQs. In this panel, we did not fit an exponential distribution.}
        \label{evolraw2}
\end{figure}

Remarkably, a close observation of seismicity (Fig.~\ref{evolraw}a) suggests that the number of occurrences in class S1 and S2 became null about 30 days before the Tohoku-oki EQ. A similar anomaly is also observed for V1 (Fig.~\ref{evolraw}b). To statistically evaluate the anomaly of such quiescence, we focus on an inter-event time distribution within each class of LFE and volcanic tremors. To this end, we reformulated the inter-time of consecutive events for a given class label. We can expect that an event in each class occurs following Poisson process with sufficient time from the previous occurrence (Fig.~\ref{evolraw2}a, b; but this is not the case for conventional EQs, Fig.~\ref{evolraw2}c). Hence, we evaluate p-values of the length of null seismicity from the Tohoku-oki EQ (Table~\ref{GaussEstimation}), based on an exponential distribution for each class. We arbitrarily truncated the fitted data to one day. For LFEs, the p-values in classes S1 and S2 were significant at 0.05 with Bonferroni correction \cite{rice2006mathematical}. On the other hand,  for volcanic tremors it was significant only for class V1. In summary, these results suggest that classes S1, S2 and V1 may work as a short-term precursor for the Tohoku-oki EQ. In contrast, no anomalous quiescence was observed for conventional EQs (Fig.~\ref{evolraw}c, Fig.~\ref{evolraw2}c). 

\begin{figure}[]
     \begin{center}
       \includegraphics[scale=0.7, trim=0mm 0mm 0mm 0mm]{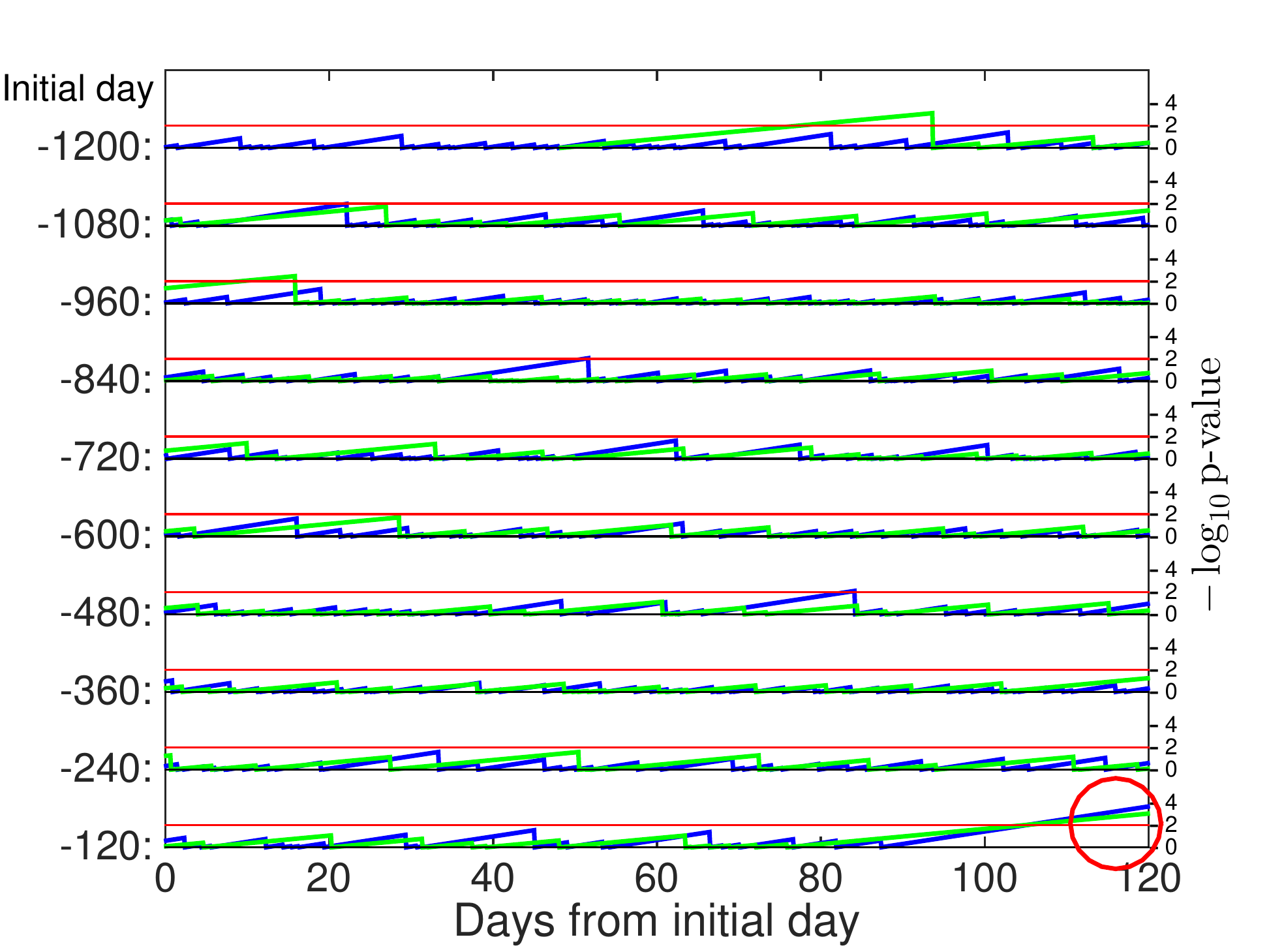}  
                \end{center}
    \vspace{0mm}
        \caption{Evolution of p-values of inter-event time for class S1 (blue) and V1 (green), respectively. 
        X-axis denotes time (day) and y-axis negative logarithm of p-values with base 10. Here,  time is evaluated as days from the Tohoku-oki EQ. Since volcanic tremors of class V1 are observed only after -1152 days, we consider the time span from -1200 days to 0 from the Tohoku-oki EQ, which is split into 10 sub-time spans. Each row in the plot denotes a particular sub-time span in which the initial day is shown at the left. 
     P-values are evaluated using parameters in a fitted exponential distributions as detailed in the caption of  Fig.\ref{evolraw2}, discarding inter-event times smaller than the cutoff inter-time that defines S1 and V1, respectively. 
        The horizontal red line denotes the significance level 0.01 (the corresponding negative logarithm is 2).  
        P-values of both class S1 and class V1 take four just before the Tohoku-oki EQ (red circle).}
        \label{precursor}
\end{figure}

Further, we examined how significant these short-term anomalies are, compared with other periods of time. The inter-event time for both class S1 and V1 seems to follow an exponential distribution (Fig.~\ref{evolraw2}a, b) except for the tails.  However, the observed values of long inter-event times before the Tohoku-oki EQ do not seem to lie in the range of the corresponding exponential distribution. This suggests that such anomalous long quiescence just before the Tohoku-oki EQ may have been induced by mechanisms different from those at other periods of time. Note that such long inter-event time is not unique to the period just before the Tohoku-oki EQ. For class S1, a long inter-event time (longer than 30 days) was observed in mid August 2001 and early April 2005, and for class V1 in mid January 2008. Within two months of the onset of these quiescence, no earthquake larger than magnitude 7 occurred in northern Japan. However, the simultaneous quiescence of both S1 and V1 is unique to the period between Day -32 and Day 0 (the day of the Tohoku-oki EQ). To clarify this, we evaluated p-values of inter-event times in continuous time space,  which is defined as the time from the latest event of class S1 (or, class V1), using the fitted exponential distribution. We found that observed p-values as low as $10^{-3}$ simultaneously for both classes S1 and V1 are unique to the Tohoku-oki EQ (Fig.~\ref{precursor}). Simple calculation based on exponential distributions fitted to these classes suggests that such simultaneous quiescence of S1 and V1 is quite a rare event that could occur only once in 1300 years if we assume that events of these classes occur independently (for S1 once in 1/(0.27*0.00014*365) =72 years; for V1 once in 1/(0.16*0.0009*365)=19 years). Certainly, this estimate is rather naive, because there could be some correlation of inter-event times between S1 and V1, but it is beyond the scope of the present paper to evaluate possible dependence between S1 and V1. 

\section*{Discussion}
\begin{figure}[]
     \begin{center}
      \includegraphics[scale=0.4, trim=40mm 0mm 0mm 0mm] {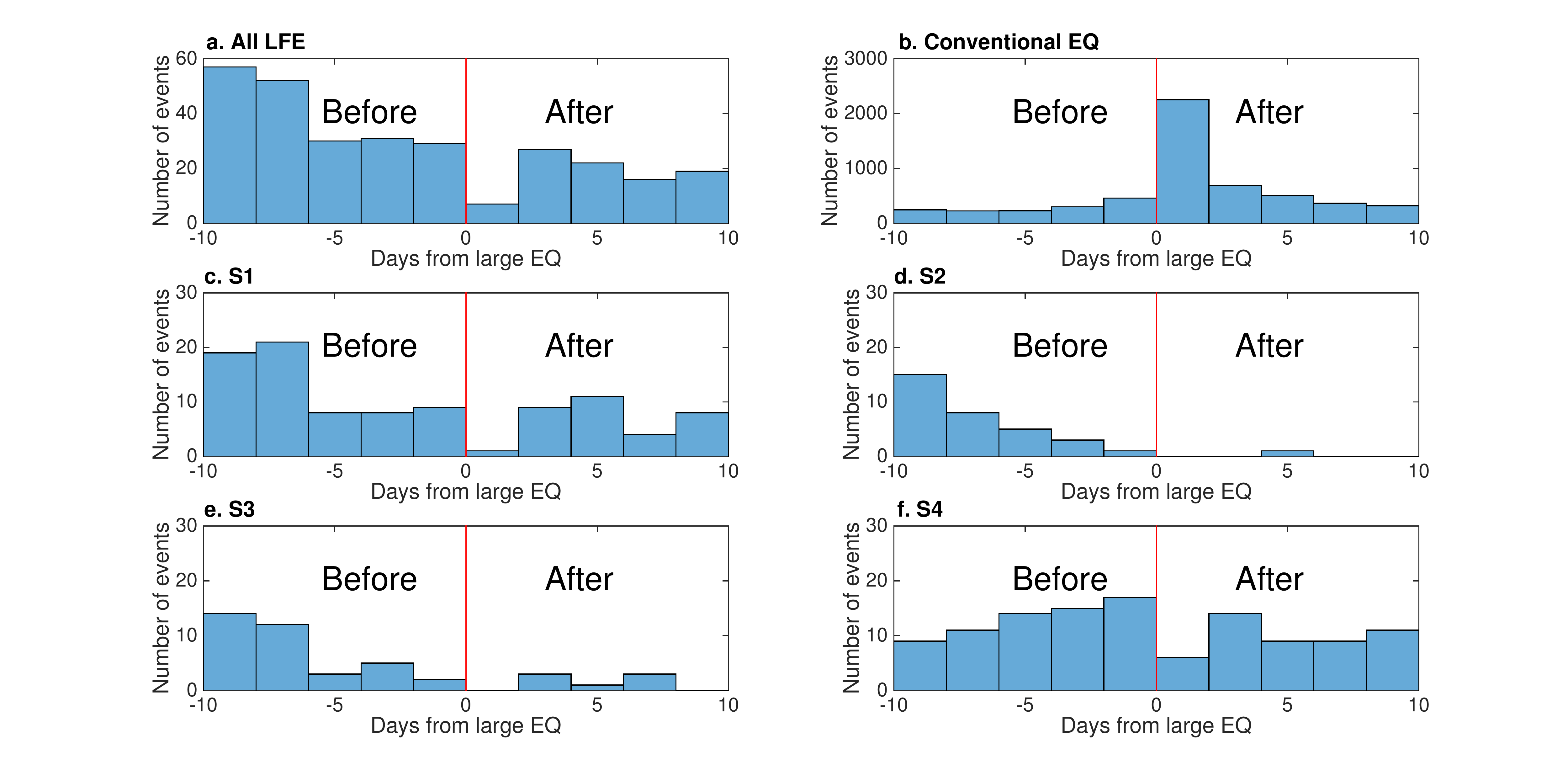} 
                  \end{center}
     \caption{The number of events before and after 61 large earthquakes with $\mbox{M}_w\geq5$. Panels a-f are for
     all LFE, conventional EQ, class S1, Class S2, Class S3 and Class 4, respectively. In this analysis, 
     we focussed on the time period between Oct.1, 2001 and Feb.11, 2011 (30 days before Tohoku-oki EQ), restricting areas: 
     Latitude between 39\degree N and 41\degree N; longitude between 139\degree E and 145\degree E. For conventional EQ, we counted the number of events with $\mbox{M}_w\geq 2$.  P-values of chi-square test for difference of number of events 
     between before and after the large earthquakes: $1.2 \times 10^{-10}$ 
     for all LFE; $4.6 \times 10^{-274}$ for conventional EQ; 
      $6.6 \times 10^{-4}$ for class S1; $3.5 \times 10^{-8}$ for class S2; $5.1 \times 10^{-6}$ for class S3; 
     0.07 for class S4. Except class S4, the difference of seismicity between before and after large earthquakes are significant at 0.05 level. However, seismicity is larger after large earthquakes in case of conventional EQ, while
     it is larger before large earthquakes in case of all LEF, classes S1, S2, and S3. }
\label{FigS2}
\end{figure}

First, we discuss interpretations of both LFEs and conventional EQs in terms of fitting different types of probabilistic distributions. Let us assume that the occurrence of LFEs follows a non-homogeneous (compound) Poisson distribution \cite{daley2007introduction, de2016statistical} conditional on the rate parameter $\theta$, the distribution of inter-time $x$ is given by an exponential distribution $\theta \exp(-\theta x)$. Also, denoting weight for $\theta$ as $g(\theta)$ (probability density function), the marginal distribution of the inter-time can be in general expressed by
\begin{eqnarray}
  f(x) = \int_{0}^{\infty} \theta \exp(-\theta x) g(\theta) d\theta.
  \label{inhomoT}
\end{eqnarray}
The probability density $g(\theta)$ can be modeled as a function of time $t$ in the framework of Ohmori's law \cite{bottiglieri2010multiple}. However, in the case of LFEs, we have little evidence that the seismicity of LFEs follows Ohmori's law (Fig.~\ref{FigS2}). For simplicity, we do not parameterize $\theta$ with time $t$. 

\begin{figure}[]
     \begin{center}
 \includegraphics[scale=0.5, trim=90mm 0mm 0mm 0mm]{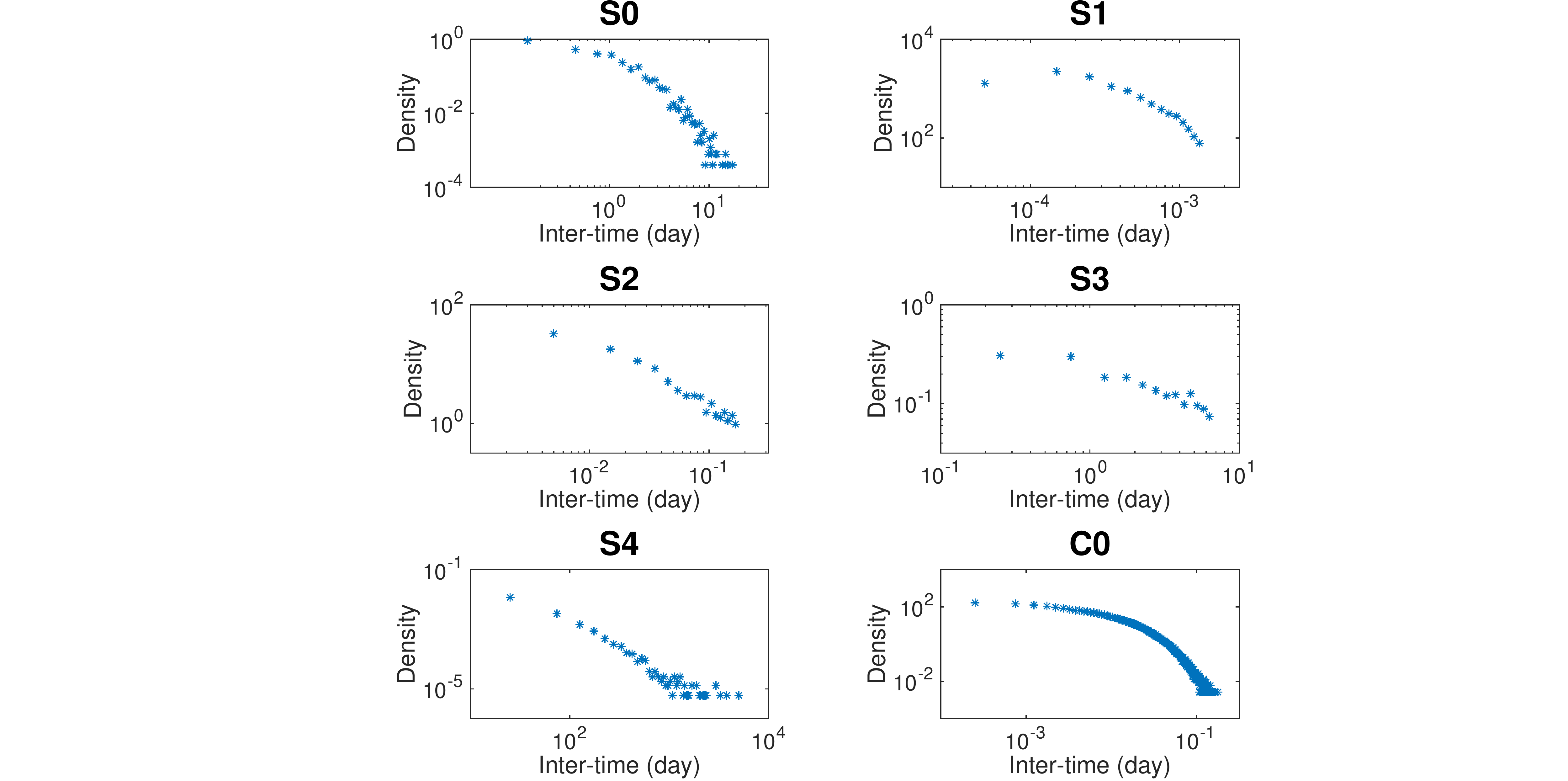}
                \end{center}
    \vspace{0mm}
        \caption{Log-log plots of inter-time and probability density for 
      classes S0-S4 and C0. }
        \label{FigS3}
\end{figure}

\begin{figure}[]
     \begin{center} 
 \includegraphics[scale=0.5, trim=90mm 0mm 0mm 0mm]{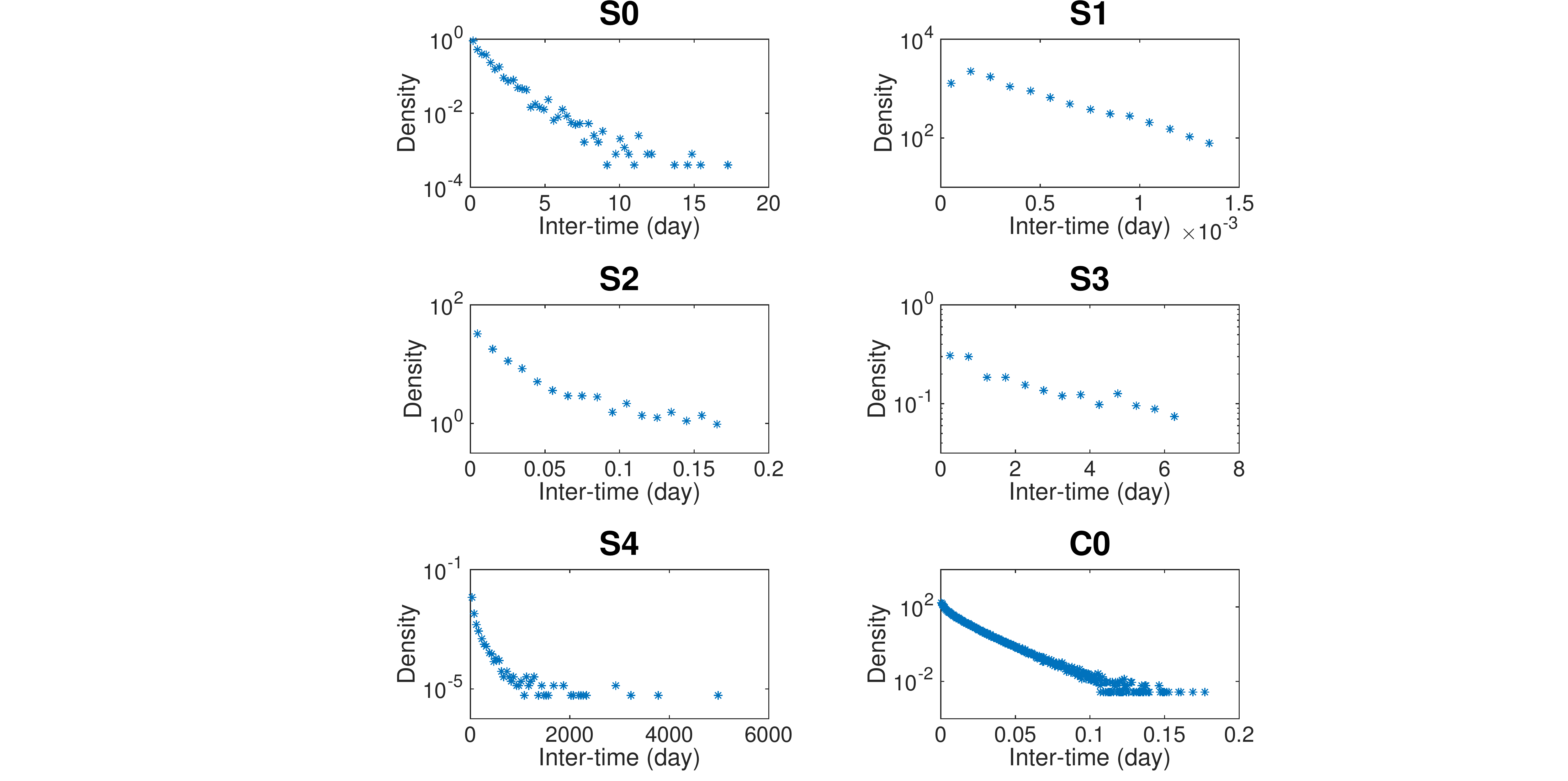}
                \end{center}
    \vspace{0mm}
        \caption{Semilog plots of inter-time and probability density 
        for classes S0-S4 and C0.}
        \label{FigS4}
\end{figure}

\begin{figure}[]
     \begin{center}
                      \includegraphics[scale=0.4, trim=0mm 0mm 0mm 0mm]{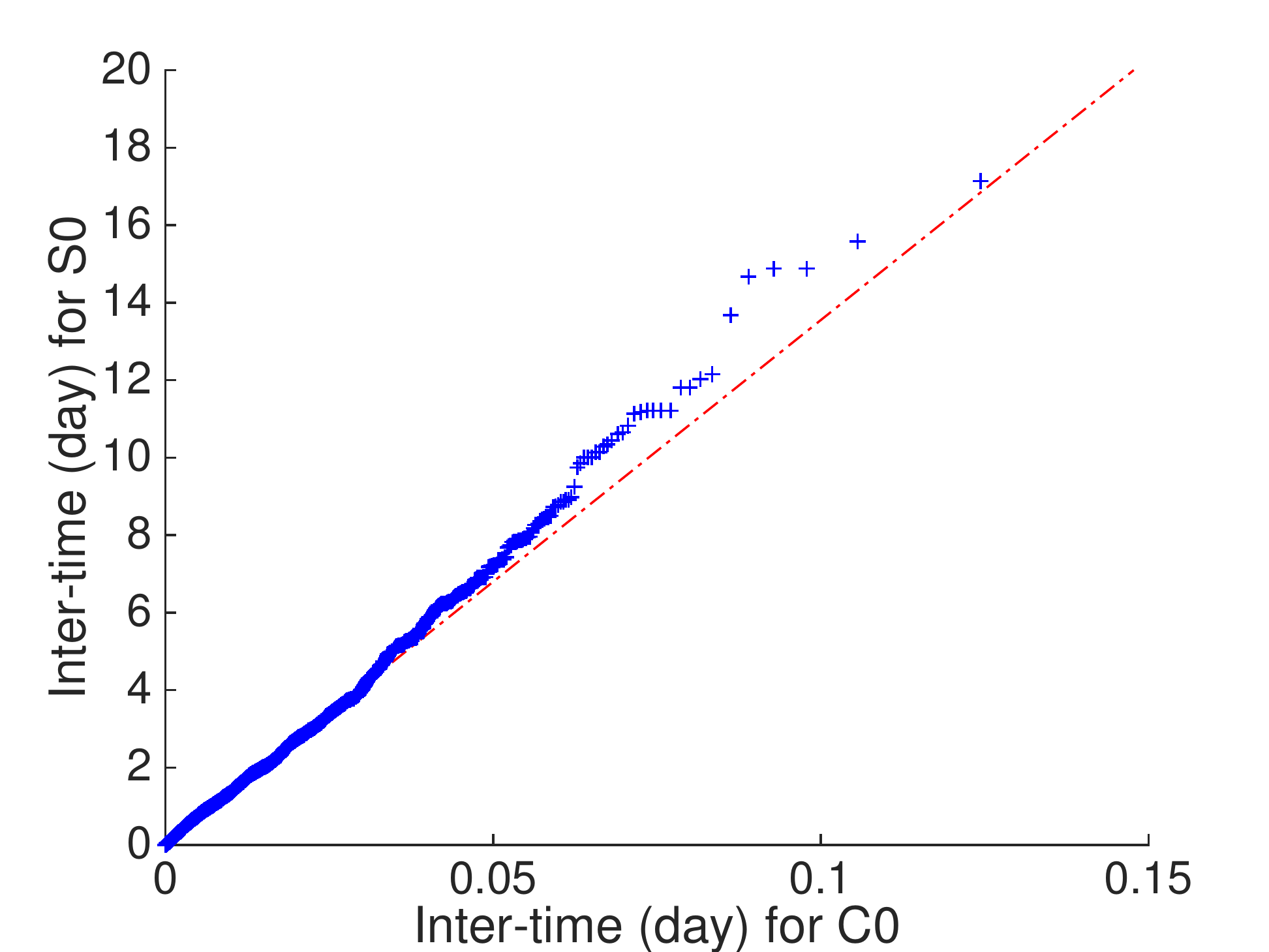}      
                \end{center}
    \vspace{0mm}
        \caption{The qq-plot \cite{rice2006mathematical} of inter-time distributions between 
        class C0 and S0.  Each dot denotes a particular quantile for both distributions of class C0 and class S0. If the shape of density distributions is the same between two, these dots are supposed to lie in the red line. 
        }
       \label{FigS5}
\end{figure}

\begin{figure}[ht!]
     \begin{center}
     \includegraphics[scale=0.4, trim=50mm 50mm 0mm 0mm]{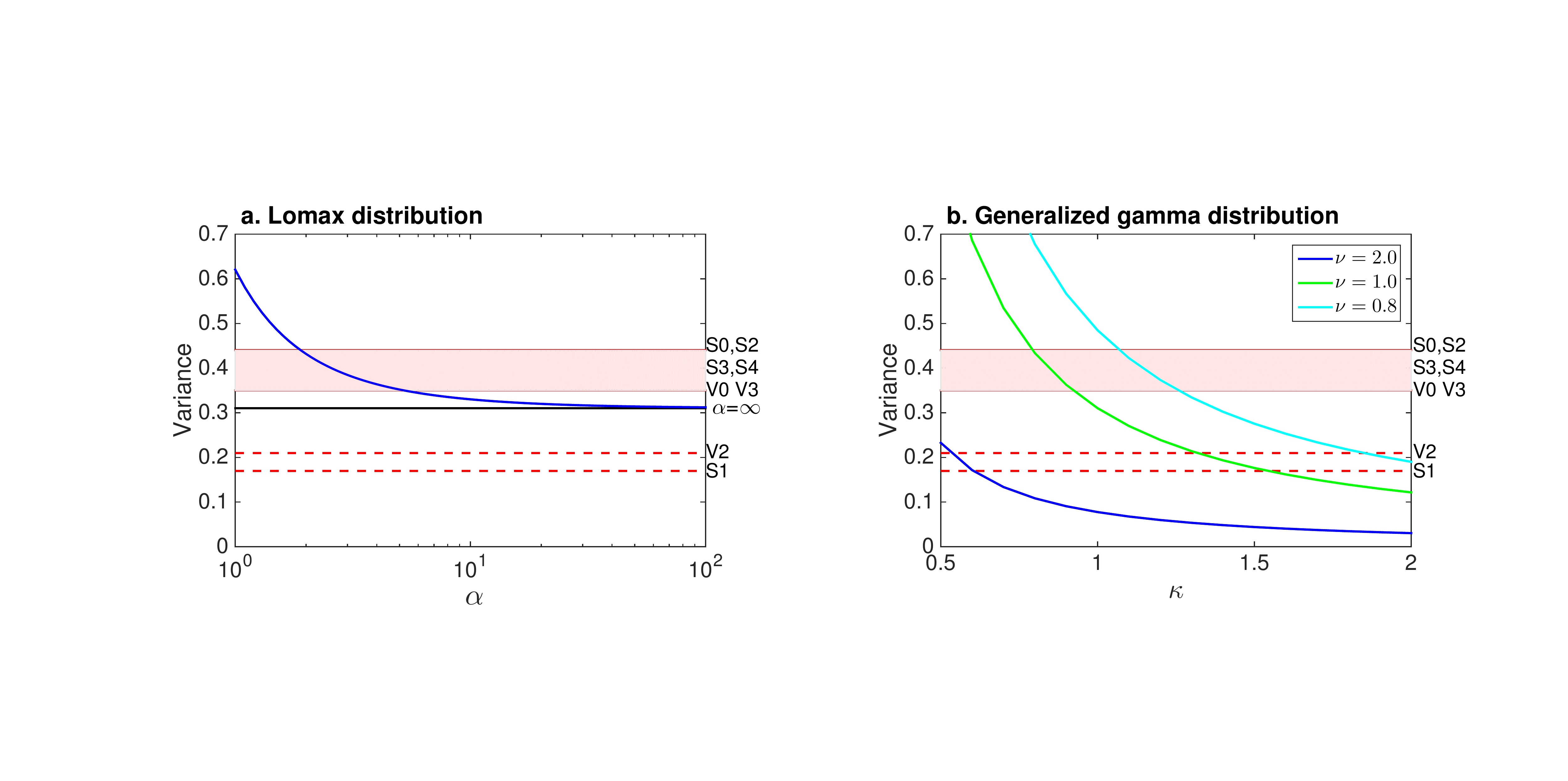}
                \end{center}
    \vspace{0mm}
        \caption{Theoretical values of variances for the logarithm of inter-time as a function of a given parameter to a specific distriubtion. 
        Panel a:  For Lomax distribution. The x-axis denotes the parameter $\alpha$ of Eq.(3) in the main text, while the y-axis the variance of the logarithm of inter-time. The red area denotes the range of estimated variances of class S0, S2, S3, S4, V0, and V3 (Table 1 in the main text), while the dashed line denotes the estimate variances of class V2 and 
        S1. The black line denotes the asymptotic value of variances as $\alpha \rightarrow \infty$. Panel b: For a generalized gamma distribution. The x-axis denotes $\kappa$ in Eq.(4) in the main text. We manipulated the parameter 
$\nu$ to 0.8, 1 and 2.  It can be shown that for fixed value $\nu$, the variance converges to 0 as $\kappa \rightarrow \infty$.}
    \label{FigS6}
\end{figure}

In the case of LFEs, our empirical data suggest that $f(x)$ may follow a power law for S2-S4, but not for S1 (Fig.~\ref{FigS3}, Fig.~\ref{FigS4}). For classes C0 and S0, we carry out more vigorous analysis , because full data are available for these classes. Classes C0 and S0 have similar variances in the logarithm of inter-time distribution (Table~\ref{GaussEstimation}), but it is observed that these two classes may follow different generative 
models (Fig.~\ref{FigS5}). We examine model-fitting to inter-time in these classes by two types of distributions: Lomax distribution (or, Pareto type II distribution) \cite{giles2013bias}
 \begin{eqnarray}
f(x) = \frac{\alpha \beta^\alpha}{(x+\beta)^{\alpha+1}}
\label{lomaxx}
 \end{eqnarray}
and a generalized gamma distribution \cite{corral2004long}
 \begin{eqnarray}
 f(x) = \frac{\nu}{\sigma^{\nu \kappa}\Gamma(\kappa)} x^{\nu \kappa-1}\exp \{-(x/\sigma)^\nu\},
 \label{gamgeneral}
  \end{eqnarray}
where all relevant parameters are positive. For classes C0 and S0, a BIC-based model selection suggests a generalized gamma distribution. The exponent of $x$ in Eq.(\ref{gamgeneral}), i.e., $\nu*\kappa-1$, takes values between -0.1 and 0.0 for both classes, which suggests that the tail of the distribution decays exponentially (Table~\ref{gampareto}). Regarding classes S1, S2, S3, and S4,  full data are not available. The aforementioned model selection is not straightforward, because the method of fitting a truncated generalized gamma distribution is not well established. Hence, we consider an approach of matching variances of the logarithm of inter-time. It can be shown that the variance of $\log_{10} x$ for a Lomax distribution in Eq.(\ref{lomaxx}) is analytically given by $(\psi '(\alpha)+\pi^2/6)/(\log 10)^2$ where $\psi '(\alpha)= \sum_{k=0}^{\infty} 1/(k+\alpha)^2$ is the first derivative of the digamma function with respect to $\alpha$. Note that the variance is not related to $\beta$ (because the variance of the logarithm of  inter-time is invariant of scales of inter-time). Since $\psi '(\alpha)$ monotonously decreases, so does the variance
with respect to $\alpha$ (Fig.~\ref{FigS6}), which converges to $(\pi ^2/6)/(\log 10)^2 \approx 0.310$ as $\alpha \rightarrow \infty$. It can be shown that a homogenous Poisson distribution corresponds the case of $\alpha \rightarrow \infty$. This analytical result can explain the variances of the logarithm of inter-time of S0, S2, S3, and S4 (also V0 and V3), suggesting the range of values of $\alpha$ between 1.9 and 5.2. In contrast with S0, S2, S3, and S4, 
the variance for class S1 (as well as V2) takes a much smaller value (0.17, Table~\ref{GaussEstimation})  than the minimum value (0.310) suggested by a Lomax distribution. This implies that the inter-time of class S1 (as well as V2) does not follow a Lomax distribution. Next, we consider a generalized gamma distribution. It can be shown that the variance of the logarithm of inter-time is given by $\psi '(\kappa)/(\nu\log 10)^2$, which is in similar form to that of a Lowmax, but without the additive term $\pi^2/6$. It can be shown that the variance of the logarithm of inter-time can take those estimated variances of all classes including S1 and V2 by tuning $\nu$ and $\kappa$ (Fig.~\ref{FigS6}). These results suggest that inter-time in class S1 should follow a generalized gamma distribution while for classes S2, S3, S4 (also V0 and V3) we cannot draw a definitive conclusion on model selection. 

\begin{figure}[ht!]
     \begin{center}
      \includegraphics[scale=0.6, trim=0mm 0mm 0mm 0mm]{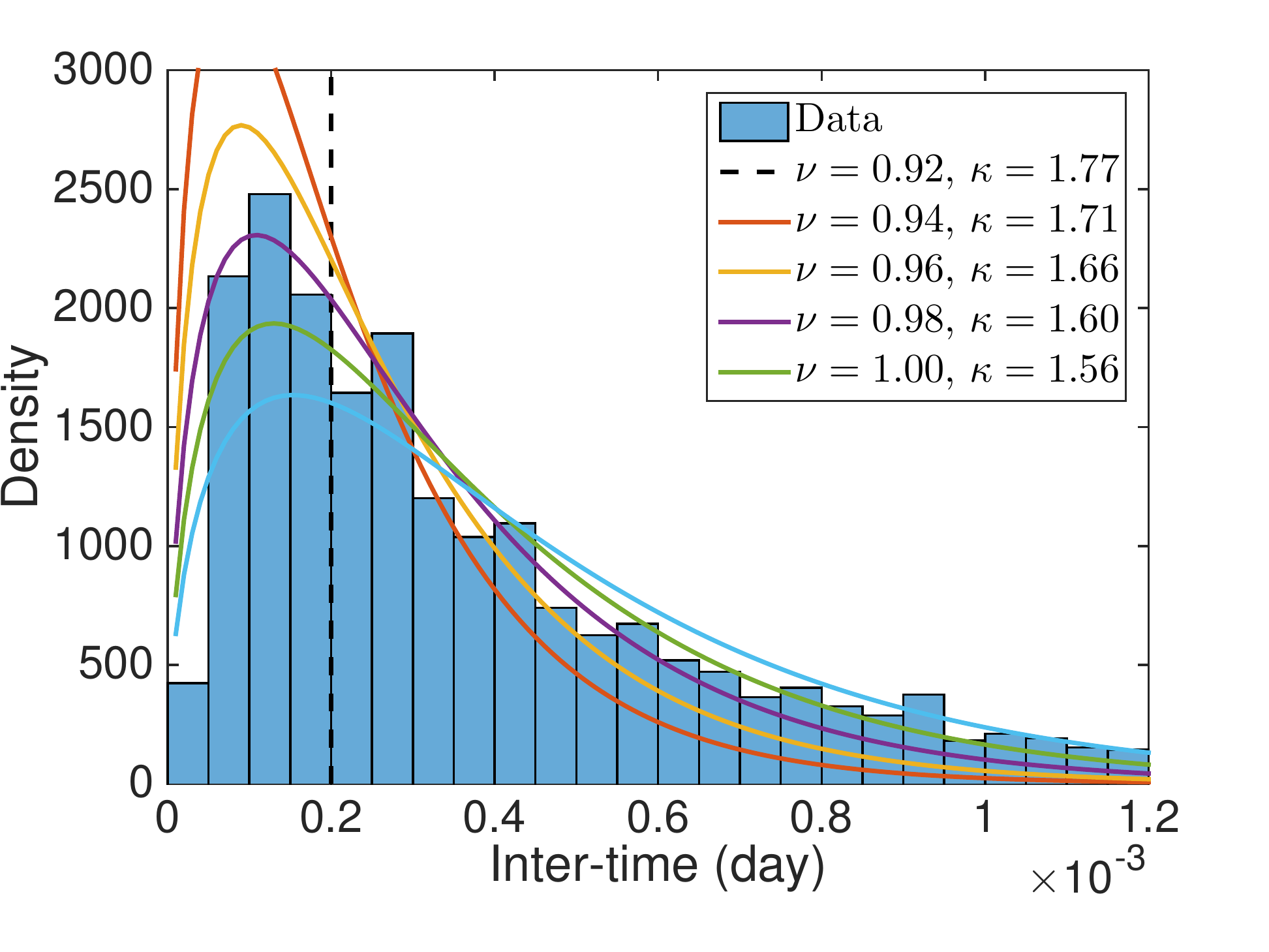}  
                \end{center}
    \vspace{0mm}
    \caption{Density functions of inter-time of class S1. We fitted a generalized gamma distribution to the data, imposing the constraint that the mean and the variance of the logarithm of inter-time should match those estimated values  of class S1, -3.51 and 0.172, respectively. With this constraint,  the triple of parameters $\sigma$, $\nu$, and $\kappa$ in a generalized gamma distribution has one degree of freedom. We manipulated $\nu$ while evaluating $\sigma$ and  $\kappa$ using the constraint of the mean and the variance. The dotted black line denotes the lower cutoff value $0.2\times 10^{-3}$ of inter-time, by which we truncated the data for re-fitting a generalized gamma distribution. 
    }
       \label{FigS7}
\end{figure}

For class S1, we further estimate parameters in a generalized gamma distribution. Matching means, given by $(\log \sigma + \psi (\kappa))/\log 10$, and variances to the data reduces the degree of freedom of the parameters $\sigma$, $\nu$ and $\kappa$ to one. Using this constraint, we estimate these parameters based on the principle of maximization of likelihood. The results of optimization suggest that $\nu \approx 0.98$  and $\kappa \approx 1.6$ (Table~\ref{gampareto}; Fig.~\ref{FigS7}). The value of $\nu$ close to one suggests that a gamma distribution may well fit the data. This result is similar to the case of conventional EQs with the lower cutoff magnitude ranging from 5 to 6.5 in global seismic catalogs \cite{corral2004long}, and the case of conventional EQs in northern Japan before the Tohoku-oki EQ  with the lower cutoff magnitude 4 (\mysingleq{C0M4b2011} in Table~\ref{gampareto}). Note that the parameter $\nu$ after the Tohoku-oki EQ considerably changed from the parameter $\nu$ before the Tohoku-oki EQ as is seen from the difference of $\nu$ between \mysingleq{C0M4b2011} and \mysingleq{C0M4} (conventional EQ with cutoff magnitude 4 in the whole period in our study). In case of conventional EQs in global seismic catalogs, the value of $\kappa$ is estimated to $\kappa \approx 0.67$; hence, the exponent of inter-time in Eq.(\ref{gamgeneral}) becomes $\nu*\kappa -1 = -0.33~(<0)$, suggesting high frequency of earthquakes in a short period of time. On the other hand, in class S1, the exponent becomes $0.54~(>0)$, suggesting the repulsive nature of occurrences between two events. This result reveals a fundamental difference in occurrence of events: In  conventional EQs, soon after a pre-event, it is more likely that a post-event will occur. In the case of class S1, soon after a pre-event, there is a quiescent time before a post-event. One may wonder whether the inferred quiescent period for a post-event in class S1 may be attributed to a technical problem of detectability of consecutive events in a short period of time. We examined this issue to explicitly remove the problem of detectability. In Fig.~\ref{FigS7}, it is observed that events with inter-times between  $0.1 \times 10^{-3}$ day (8 seconds) and $0.15 \times 10^{-3}$ day (12 seconds) are most likely to occur.  Now, we fit a left- and right-truncated generalized gamma distribution to the inter-time of class S1. We keep the same upper cutoff as before, but we set the lower cutoff to $0.2 \times 10^{-3}$ (17.2 seconds), which has the effect of removing inter-times less than the lower cutoff (i.e., removing four bins from the left in Fig.~\ref{FigS7}). As a result of fitting, we obtained $\nu=1.02$, $\kappa=1.48$ and $\nu * \kappa -1 \approx 0.51$, which are similar to the case without a lower cutoff. This confirms the quiescent period for a post-event in class S1. 

Second, we discuss interpretations of class S1 as a precursor to the Tohoku-oki EQ. Our analysis suggests that seismic quiescence of class S1 showed up at least three months before the Tohoku-oki EQ, and that subsequently this type of LFE completely disappeared 30 days before the Tohoku-oki EQ (Table~\ref{GaussEstimation}). The timing of onset of seismic quiescence is consistent with the timing of the observed abnormal change of level and temperature of groundwater in Goya-onsen, 155 km northwest of the epicenter of the Tohoku-oki EQ.  At Goya-onsen, an anomalous drop of water level and temperature began  three months before the Tohoku-oki EQ \cite{orihara2014preseismic}. Similarly, an anomalous increase of Radon concentration was observed in the Izu Peninsula, 460 km southwest of the epicenter, three months before the Tohoku-oki EQ \cite{tsunomori2014anomalous}. Furthermore, the timing of the complete disappearance of class S1 is consistent with the timing of onset of the presumed nucleation process near the epicenter \cite{kato2012propagation}. In addition, our observation of quiescence of class S1 can be contrasted with the observed quiescence of conventional EQs, which began 23 years before the Tohoku-oki EQ \cite{katsumata2011long}. Quiescence of conventional EQs may work as a long-term indicator, whereas class S1 may play a key role as an immediate harbinger of a large earthquake. 

Now, we explore a possible geodetic interpretation for the observed phenomenon in LFE. In general,  seismicity reflects accumulation of strains \cite{ogata2005detection}. Our results suggest that across much of Tohoku, the accumulation of strain was reduced prior to the Tohoku-oki EQ. A possible explanation for this phenomenon may be reduced movement of the North American plate in which the Tohoku region lies. On the other hand,  during the same period of time, it is reported that a nucleation process took place near the epicenter \cite{kato2012propagation}, which suggests an increment of strain. It is worth noting that the nucleation process began 30 days before the Tohoku-oki EQ, which coincides with the onset of the complete quiescence of class S1. These seemingly contradictory phenomena may be explained by the asperity model, which assumes strong coupling of some areas of the plate interface in subduction zones \cite{lay1981asperity}. Several studies suggest the existence of asperities off-shore of Tohoku \cite{yamanaka2004asperity, johnson2012challenging}, which is located west of the epicenter. These asperities lie geographically between the area of LFEs in the present study and the area of the nucleation process \cite{kato2012propagation}. Based on the asperity model, we offer a possible explanation as follows. Long before the Tohoku-oki EQ, the Pacific plate continued to push the North American plate to the west, but 30 days before, the asperities became strongly coupled. Hence, the strain accumulated east of the asperities, i.e., near the epicenter. On the other hand, west of the asperities, the rate of accumulation of strain became reduced. 
% * <steven.aird@oist.jp> 2018-07-10T07:06:11.420Z:
% 
% > show
% Tomoki, this word is wrong, but I do not know what you wanted to say.
% 
% ^ <steven.aird@oist.jp> 2018-07-10T07:06:39.132Z.

Finally, we discuss limitations of the present study. First, to analyze seismicity from the LFE data, missing data could be a statistical issue, potentially biasing the data analysis. In general, LFEs of low magnitude are less likely to be captured by the network of seismic meters than those of large magnitude. In the present study, we focused on the inter-time between two LFE events, assuming that inter-time and magnitude of LFE are independent. Second, we assume that data availability of LFE in the JMA catalog did not change during our study.  There are two factors that may challenge this assumption. One factor is the possible improvement of detective capability of LFEs owing to technical advancement of seismic meters. The major change in this period is that  some stations, including Sendai, located in the Tohoku region started to use an F-net seismic network\cite{okada2004recent} to upgrade their detective capability for earthquakes on Oct.1, 2001 \cite{JMA2}. Also,  the weighting function for hypocenter location program was upgraded on the same day. Taking account of this change of detective capability, we used the data for LFEs and conventional EQs after October 1, 2001. We suppose improvement of further possible detective capability has little influence on the results in the present study. The other factor is effect of the Tohoku-oki EQ: A large number of aftershocks might have masked LFEs in seismic meters. If this is true, we expect that not only LFEs, but also conventional EQs with small magnitude would be masked. To examine this, we evaluated a cutoff value of magnitude that ensures the Gutenberg Richter law for conventional EQs in the region of our study. It is observed that the cutoff magnitude was $M_c \approx 0.5$ before the Tohoku-oki EQ, but that it became larger thereafter ($M_c \approx 1.2$). However, after 9 months the cutoff magnitude returned to its level prior to the Tohoku-oki EQ. This  observation suggests that we may not have to take into account the masking effect at least 9 months after the Tohoku-oki EQ. In Fig.~\ref{fig4}c, low activity of class S1 is still observed 9 months (270 days) after the Tohoku-oki EQ. We believe that the long-term anomaly in S1 would be observable even if we discarded the period of aftershocks. Importantly, our analysis of the long-term anomaly based on Z-value does not use the data after the Tohoku-oki EQ; hence, the results on the long-term anomaly in Z-value are intact for the effect of the Tohoku-oki EQ. Similarly, regarding the short-term anomaly, our analysis does not use the data after the Tohoku-oki EQ, either. Hence, the results in the short-term anomaly are also intact for the effect of the Tohoku-oki EQ. 
 
\section*{Methods}
In this section, we provide details on data, pre-processing, cluster analysis, ROC curve analysis, and Z-values, which were employed to yield the results in the present paper. 

\subsection*{Data}
All data in the preset study, including LFEs, conventional EQs, and volcanic tremors, were obtained from JMA catalogs (Japan Meteorological Agency, \url{https://www.jma.go.jp/jma/indexe.html}). We used 8263 LFEs that are flagged in the seismic catalog in the period from October 1, 2001 to March 31, 2016 (i.e., -3448 to 1847 days from the Tohoku-oki EQ). We consider LFE only after October 1, 2001, because detective capability of LFE by JMA considerably improved after this date. For conventional EQs, we focused on the same time period. For volcanic tremors, we considered volcanos with latitudes between $37.5\degree$N and $42\degree$N covering a wide area of northern Japan, and with more than 100 volcanic tremors. Six volcanoes met our criteria, but in volcano Zaozan ($38\degree08'37''$N, $140\degree26'24''$E), the first volcanic tremor was not recorded before 300 days from the Tohoku-oki EQ. Due to this limited time period, we excluded this volcano for analysis. As a result, we included the following five volcanos in our analysis: Esan ($41\degree48'17''$N, $141\degree09'58''$E), 
Iwatesan ($39\degree51'09''$N, $141\degree00'04''$E),
Azumayama ($37\degree44'07''$N, $140\degree14'40''$E), 
Adatarayama ($37\degree37'59''$N, $140\degree16'59''$E) and 
Bandaisan ($37\degree36'04''$N, $140\degree04'20''$E) \cite{JMA}
with observed volcanic tremors 183, 1306, 2733, 139 and 700, respectively. The time of observed volcanic tremors was: from -1165 day to 1108 day for Esan; from -1164 day to 1107 day for Iwatesan; from -1132 day to 1094 day for Azumayama; from -1123 day to 1028 day for Adatarayama; from -1161 day to 1112 day for Bandaisan. Here, we set the origin time of the Tohoku-oki EQ to 0 day. 

\subsection*{Pre-processing}
Given a time series of EQ events, the inter-time of two events is simply defined as the time elapsed between two consecutive events. Here, we further extend this definition, taking into account the proximity of two events, as follows. First,  the time of event occurrence is sorted in ascending order, $t_1 < t_2 < \ldots < t_N$, where $t_i$ denotes the time (day) of occurrence of $i$th event; $N$ sample size. Let $\Delta_{i, j}$ be the difference of time between events $i$ and $j$ ($i<j$), defined by $ \Delta_{i, j} = t_j - t_i$. Denoting  as $d_{i, j}$ the distance (km) between epicenters of events $i$ and $j$, we define the inter-time constrained by $d_{i, j}$ as follows:
\begin{eqnarray*}
 \Delta_{i, j}' (d_{min}, d_{max})=\Delta_{i, j} \mathbb{I}(d_{i, j}>d_{min}) \mathbb{I}(d_{i, j}<d_{max}),
\end{eqnarray*}
where $\mathbb{I}(a)$ is an indicator function. This function simply sets inter-time to zero if the distance between events $i$ and $j$ is less than $d_{min}$ or larger than $d_{max}$, thus degenerating inter-time in such cases.  Using this quantity, we generate a vector of inter-time  denoted by $\bb{\Delta} (d_{min}, d_{max})$, which consists of $N$ elements $\Delta_{i}$ for $i$th event as follows:
\begin{eqnarray}
  \Delta_i (d_{min}, d_{max})&=& \underset{i<j, \Delta_{i, j}' >0}{\mbox{min}} ~  \Delta_{i, j}' (d_{min}, d_{max}).
  \label{difintertime}
\end{eqnarray}
In a nutshell, $\bb{\Delta}(d_{min}, d_{max})$ represents a collection of inter-time in a specific range of distance between $d_{min}$ and $d_{max}$, where we allocate the inter-time to the pre-event, rather than the post-event. 

\begin{figure}[ht!]
     \begin{center}
 \includegraphics[scale=0.5, trim=80mm 80mm 0mm 0mm]{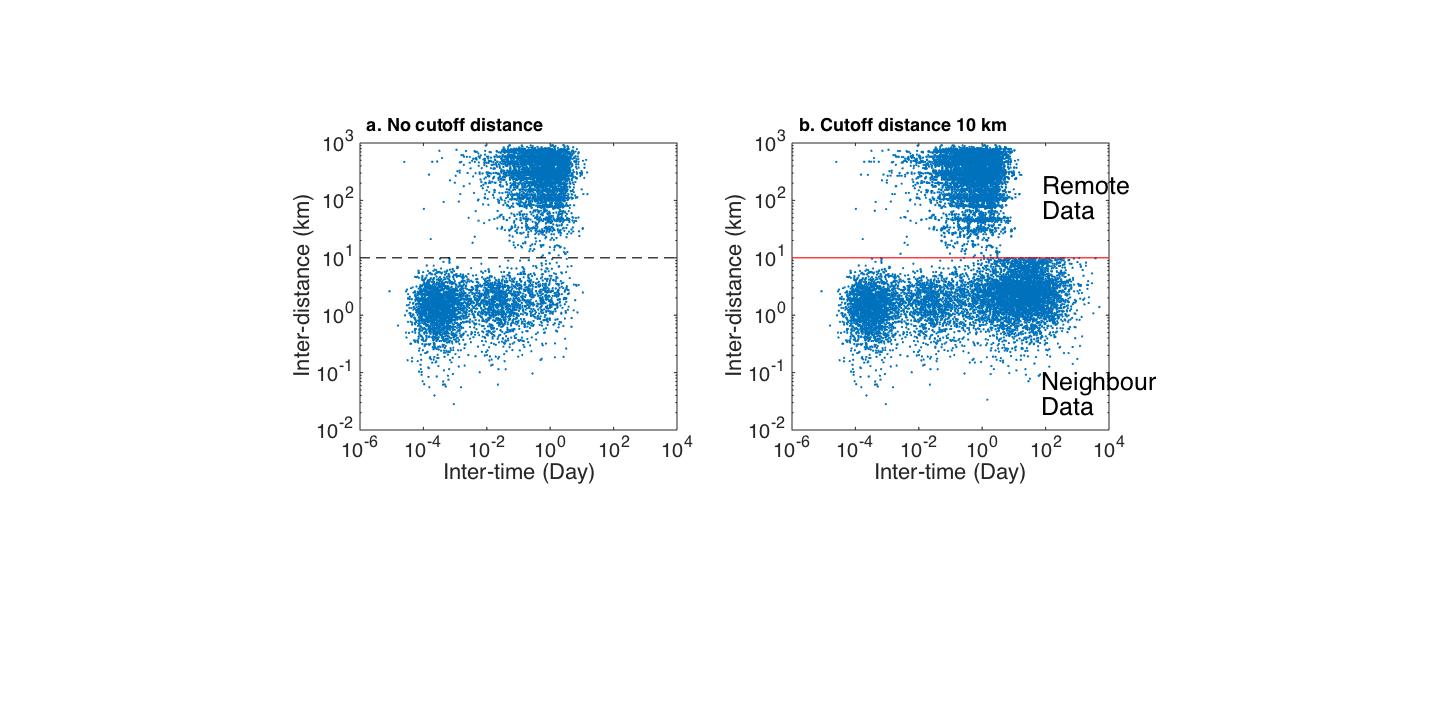} 
    \end{center}
        \caption{Distribution of LFE as a function of inter-time and inter-distance. 
        Panel a:  No inter-distance cutoff (dataset $\bb{\Delta} (0, \infty)$).   
        Panel b: With inter-distance cutoff 10 km. Dataset for remote pairs of events  ($\bb{\Delta} (10, \infty)$) in the upper part, and for neighboring paris ($\bb{\Delta} (0, 10)$) in the lower part. The red line (inter-time distance 10 km) denotes the boundary between two datasets.}
        \label{FigS1}
\end{figure}

With these notations, the dataset $\bb{\Delta}(0, \infty)$ represents a collection of inter-times without any constraint of distance. In our data, the distribution of this type of inter-time is displayed in Panel a of Fig.~\ref{FigS1}. Remarkably, it is observed that the distribution of inter-time differs between the upper part and the lower part, which are separated by the line of distance 10 km. This suggests that there are two heterogenous groups of inter-time, characterized by the inter-distance between consecutive events. 

For a better understanding of the underlying mechanism of LFEs, it would be useful to analyze these groups separately. Note, however, that the upper and the lower part are closely intertwined in this dataset. To minimize such interactions, we set the cutoff point to 10 km to generate two datasets: $\bb{\Delta}(10, \infty)$, referred to as \mysingleq{Remote LFE}, and $\bb{\Delta}(0, 10)$ as \mysingleq{Neighbour LFE}
(Panel b of Fig.~\ref{FigS1}), both of these having the sample size $N$. From the definition in Eq.~(\ref{difintertime}), Remote LFE represents the inter-time of remote pairs of events (remote inter-time, with inter-distance larger than 10 km), while Neighbour LFE the inter-time of neighboring pairs (neighboring inter-time, with inter-distance smaller than 10 km). 

\subsection*{Cluster analysis}
To estimate the underlying distribution for a given dataset, we fitted Gaussian mixture models \cite{mclachlan2004finite}. In this model, a distribution of data, denoted as $f(x)$, is given by summation of Gaussian distributions:
\begin{eqnarray*}
  f(x) = \sum_{k=1}^{K}w_{k} \times \mbox{Gauss}(x|\mu_{k}, \sigma_{k}^2),
\end{eqnarray*}
where $K$ is the (estimated) number of classes; $w_{k}$ is weight for the $k$th component; 
$\mbox{Gauss}( \cdot | \mu, \sigma^2)$ is a Gaussian distribution for the $k$th component with mean $\mu_{k}$ and variances $\sigma_{k}^2$. Importantly, we can classify data points by allocating each data point to the most plausible component in this mixture model. Further, we manipulated the value of $K$ from one to five. To select an optimal model (the value of $K$ in the present case), the most popular criteria are the AIC (Akaike Infromation Criterion) \cite{akaike1998information} and the BIC (Bayesian Information Criterion) \cite{schwarz1978estimating}. It is well known that AIC is not consistent (i.e., the probability of identifying the true model is not necessarily one as the sample size goes to infinity), while BIC is consistent \cite{kuha2004aic}. In our dataset, the sample size is relatively large (in the order of 1000).  We accordingly used BIC for model selection in the present study. 

\subsection*{ROC curve analysis}
Suppose we have $n$ pairs of data $(x_i, y_i)$ $(1\leq i \leq n)$ where $x_i$ is numerical and $y_i$ is binary. Now, let us assume that there is an unknown association between $x_i$ and $y_i$. We considered classifying these data points using only information on $\bb{x}$ and evaluate its performance, referring to the true label $\bb{y}$. Here, our classifier was a binary classifier achieved by setting a value $x_0$: if $x_i<x_0$, we allocate $x_i$ to one group, otherwise to the other group. Our question was, "to what extent this classifier can reveal the true label?" To answer this question, we drew a ROC curve by manipulating a value of $x_0$: a ROC curve represents a graphical plot of sensitivity verse (1-specificity) where sensitivity is defined as the number of true positive/(the number of true positive + the number of false negative); specificity is the number of true negative/(the number of true negative + the number of false positive). In this plot, the horizontal axis is (1-specificity) while the vertical axis is sensitivity. AUC represents the area surrounded by the ROC curve, the horizontal axis and the vertical line that passes through (1, 1). AUC takes a value between 0 and 1. A close value of AUC to one suggests a $\bb{x}$-based classifier can yield the true label $\bb{y}$. 
%Further, we can identify an optimal value of cutoff point $x_0$. 

\subsection*{Z-value}
\mysingleq{Z-value} evaluates seismicity in a target period against a background period, which is defined as 
\begin{eqnarray*}
Z=(R_{bg}-R_{w}) /(S_{bg}/n_{bg} + S_w/n_w)^{1/2},
\end{eqnarray*} where $R_{bg}$ and $R_{w}$ are mean seismicity in the background period and the target period, respectively; $S_{bg}$ and $S_w$ are variances in the corresponding periods; $n_{bg}$ and $n_w$ the number of bins in the corresponding periods. In a nutshell,  Z-value denotes the normalized difference of seismicity between the background period and the target period. In the present paper, we divided the whole period of observations from Jan.1, 2006 to March 8, 2011 into bins of 14 days. We set the width of window ($T_w$) for a target period to 120 days with a moving step of 14 days. Note that a background period is defined as a set difference between the whole period and a target period. In this setting, we  counted the number of events, evaluating seismicity in each bin. Using seismicity in a bin, we evaluated means and variances in the target and the background periods, which led to the Z-value. 

%\bibliography{bib2.bib}
\bibliographystyle{naturemag}

\clearpage
\section*{Acknowledgments}
We thank Takahiro Hatano at the University of Tokyo for  his valuable comments and helpful suggestions. We thank Koji Tamaribuchi at Meteorological Research Institute, JMA, for clarifying the details on the completeness of the JMA catalogue. We thank Steven D. Aird at Okinawa Institute of Science and Technology Graduate University for editing this article. This study was partially supported by the Earthquake Research Institute cooperative research program No. 2016-B-07 at the University of Tokyo. 

%\section*{Author Contributions}
%TT and HS equally contributed to this paper in all aspects. 

%\%section*{Competing Interests}
%The authors declare no competing interests.

\section*{Data availability}
All data used in this study are available from JMA catalogs (Japan Meteorological Agency, \url{https://www.jma.go.jp/jma/indexe.html}).

\clearpage
% Table 1
\begin{table}
\caption{Characteristic of each class of earthquake. Weight, mean, and variances are based on Gaussian mixtures for logarithms (base 10) of inter-time. ROC cutoff is the optimal value of cutoff to 
split inter-times into two segments for ROC analysis. Quiescent period denotes the period between the origin time of the Tohoku-oki EQ and the time that the last event of corresponding class took place before the Tohoku-oki EQ. P-values were evaluated by fitting a left-truncated exponential distribution (truncated point is one day) to the data before the Tohoku-oki EQ, where the mean value of the distribution was estimated using a robust statistic, median/$\log 2$ \cite{gather1999robust}. Asterisks denote level of significance of p-values: *** $p<0.001$; ** $p<0.01$; * $p<0.05$.
 }
\begin{center}
\footnotesize
\begin{tabular}{clclrccccccl}
 \hline
 \multicolumn{2}{c}{} &   &  && &&  &\multicolumn{4}{c}{Seismicity Anomaly}\\
 \cline{9-12}
  \multicolumn{2}{c}{} &   & & & \multicolumn{3}{c}{Gaussian distribution fitted}  & \multicolumn{2}{c}{Long-term}&\multicolumn{2}{c}{Short-term}\\
   \cline{9-12}
 \multicolumn{2}{c}{} &   & Median &\multicolumn{1}{l}{Sample} & \multicolumn{3}{c}{to logarithm of inter-time}  
 &ROC&&Quiescent&\\
 \cline{6-8}
EQ Type& Data Type &  Class & Inter-time & \multicolumn{1}{l}{Size} &Weight  & Mean  &Variance &Cutoff& AUC&Period&P-value\\
\hline
& Remote & S0 & 18 hr & 8263& 1 &    -0.24 & 0.40 && &&  \\
 \cline{2-12}
& Neighbour &S1  & 24 sec & 2081 & 0.27 & -3.51&  0.17 &-76 day& 0.83& 32.7 days& $1.4\times 10^{-4~***}$\\
  LFE & & S2  & 27 min& 1336 & 0.15 & -1.63 &0.35 &249 day& 0.44 & 50.2 days& $2.4\times 10^{-3~**}$ \\
 && S3 & 2.0 day&1599 & 0.13& 0.21 & 0.42 &-53 day &0.72& 11.9 days &$2.0 \times 10^{-2~*}$ \\
 & &S4  & 35 day &3168 & 0.45 &  1.42&  0.44 &-88 day& 0.70& 1.06 days  &$5.9 \times 10^{-1}$ \\
\hline
 &Remote & V0 & 13 hr &6713 & 1 &  -0.37&  0.41 &&&&\\
 \cline{2-12}
 &Neighbour & V1  & 1.0 min & 810 & 0.13 &  $<$ -2.51 &NA &273 day&0.60& 44.1 days& $9.0\times 10^{-4~***}$\\
 Volcanic &  & V2  & 1.5 hr&2514  & 0.27 & -1.03 & 0.21&300 day&0.77&3.84 days&$1.7\times 10^{-1}$ \\
 Tremor && V3 &1.2 day& 3388 & 0.53 & 0.02 & 0.40 &300 day& 0.85& 1.12 days&$3.4\times 10^{-1}$\\
 \hline
 & Remote  & C0 & 7.0 min & 700698 & 1 &   -2.40 & 0.40 &&&&\\
 \cline{2-12}
& Neighbour  &C1  & 1.4 min &   98309& 0.12 & -3.02 &  0.35&&&&\\
  Conv. & & C2  &31 min &  127240       & 0.15 & -1.47 &  0.66&&&&\\
 EQ && C3  & 4.4 hr&  92386         & 0.16 & -1.10 & 0.99 &&&&\\
 & &C4 & 23 hr &        112692              & 0.23 &  0.32 & 1.00 &&&& \\
 & &C5 & 3.1 day &   62310              & 0.22 &  0.65 & 0.81 &&&& \\
 & &C6 & 16 day &    204456                & 0.13 &  0.99 & 0.31 &&&& \\
\hline
\end{tabular}
\end{center}
\label{GaussEstimation}
\end{table}

% Table 2
\begin{table}
\caption{Results of fitting a generalized gamma distribution. For purposes of comparison, we rescaled inter-time in each dataset by multiplying the mean rate of seismicity \cite{corral2004long}. The estimates of parameters are based on maximum likelihood estimation (MLE), or the hybrid of MLE and moment matching (MM) as described in section of Discussion. Class C0M4 denotes a class of conventional earthquakes with $\mbox{M}_w \geq 4$ of class C0, whereas class C0M4b2011 denotes conventional earthquakes that are obtained by restricting class C0M4 to before  2011 (hence, no influence of the 2011 Tohoku-oki EQ). The difference of parameters between C0M4 and C0M4b2011 suggests possible change in seismicity after the Tohoku-oki EQ. 
 As a reference, the results based on a global earthquake catalog \cite{corral2004long}  are also displayed.}
\begin{center}
\begin{tabular}{llcccc}
\hline
%\multicolumn{6}{c}{Fitted probabilistic distributions}   \\
%\cline{3-6}
& &\multicolumn{4}{c}{Parameters in generalized gamma dist.} \\
 \cline{3-6}
Class & Fitting Method & $\sigma$ & $\nu$ & $\kappa$ & $\nu*\kappa$ \\
\hline
S0 & MLE  &   0.66  & 0.76  & 1.2    & 0.94  \\
C0 & MLE &0.53  & 0.70   & 1.4 & 0.97  \\
C0M4 &MLE &    1.46  & 0.77   & 0.63 & 0.49  \\
C0M4b2011 & MLE   &  1.41  & 0.97   & 0.70 & 0.68 \\
\hline
C0M4b2011 & MLE + MM   &    1.50  & 1.03   & 0.67 & 0.69  \\
S1 & MLE + MM &  0.62   & 0.98  & 1.6 & 1.54  \\
\hline
Global EQ Catalog&   &    1.58& 0.98   & 0.68 & 0.67  \\
 &  &    $\pm$0.15 & $\pm$0.05   &  $\pm$0.05 &   \\
\hline
\end{tabular}
\end{center}
\label{gampareto}
\end{table}

% 0.6222    0.9800    1.5800
% Results of BIC
%        	   	Lomax             Generalized Gamma
% S0        		19324.41           19310.85
% C0    		-5307518            -5316587
% C0M4           -336.4877          -3644.124
% C0M4b2011 	5942.521 		5571.138
\clearpage

% Fig 1

\clearpage

% Fig2

\clearpage

% Fig3

\clearpage

% Fig4

% Fig5

% Fig6

 \clearpage

%% Adding Supplementary Information

% FigS2 (Fig7)

 % FigS3 (Fig8)

 % FigS4 (Fig9)

\clearpage
% FigS5 (Fig10)

% FigS6 (Fig11)

% FigS7(Fig12)

% FigS1 (Fig13)

\end{document}